# More than energy cost: Multiple benefits of the long Achilles tendon in human walking and running

Anthony J. Blazevich[1] and Jared R. Fletcher[2]

[1]*Centre for Human Performance, School of Medical and Health Sciences, Edith Cowan University, Australia*

[2]*Department of Health and Physical Education, Mount Royal University, Canada*

**Corresponding author:**
Anthony J. Blazevich
Centre for Human Performance, School of Medical and Health Sciences
Edith Cowan University, Joondalup 6027
Australia
Phone: +61 8 6304 5472
Email: a.blazevich@ecu.edu.au

**ABSTRACT**

Elastic strain energy is stored and released from long, distal tendons such as the Achilles during locomotion, reducing locomotor energy cost by minimising muscle shortening distance and speed, and thus activation. However, numerous additional, often unrecognised, advantages of long tendons may speculatively be of greater evolutionary advantage, including the reduced limb inertia afforded by shorter and lighter muscles (reducing proximal muscle force requirement); reduced energy dissipation during the foot-ground collision; capacity to store and reuse the muscle work done to dampen the vibrations triggered by foot-ground collisions; and attenuation of work-induced muscle damage. Cumulatively, these effects should reduce both neuromotor fatigue and sense of locomotor effort, allowing humans to choose to move at faster speeds for longer. As these benefits are greater at faster locomotor speeds, they are consistent with the hypothesis that running gaits used by our ancestors exerted substantial evolutionary pressure on Achilles tendon length.

**Speed or locomotor cost? The role of long, energy-storing tendons**

Muscle-tendon units in which relatively long, compliant tendons are arranged in series with a work-producing muscle are common in the distal regions of limbs of terrestrial, especially cursorial, animals (Alexander, 2002). In humans, the long Achilles tendon is located distal to the plantar flexor muscles that cross the ankle and plays a critical role in activities such as walking, running, and jumping. These tendons differ from more proximally located tendons, not only in their length but also their strain energy capacity (higher), failure/injury rating (higher force and strain yield points), and hysteresis (lower) or elasticity (higher) (Thorpe et al., 2015). The Achilles tendon in particular also differs markedly from the shorter, stiffer foot extensor tendon found in other primates (Vereecke et al., 2005), suggesting that it plays an important locomotor role that is distinct from that of our closest relatives and hinting at its evolutionary importance.

For high-speed movement tasks such as sprint running, jumping, and throwing, long in-series tendons help to overcome the shortening velocity-related limitations of muscular force production (Farris and Sawicki, 2012; Arnold et al., 2013; Lai et al., 2014). Such tendons can store energy that is produced relatively slowly by working muscles, or of gravity or inertia when external forces induce joint rotation whilst the muscles operate at relatively slow velocities (e.g. during ground contact in constant-speed running), and then release that energy at higher speeds as the tendon recoils later in the movement (Bobbert et al., 1986; Farris and Sawicki, 2012). This 'catapult' effect allows for the required force levels to be achieved at higher muscle-tendon shortening speeds than could be provided by muscles alone, or for greater muscle forces to be produced at a given muscle-tendon unit shortening speed (for review, see refs Alexander, 2002; Roberts, 2016). Muscle power amplification through the storage and release of elastic strain energy is thought to be substantive (at least 1.3 – 1.8-fold under varying conditions; Galantis and Woledge, 2003; Paluska and Herr, 2006; Sawicki et al., 2015) and to contribute decisively to performance of high-speed movements. A possible hypothesis, therefore, is that long distal tendons evolved specifically from the need to move at high speeds on rare occasions, such as when chasing prey, avoiding capture or, unique to humans, swinging an object or releasing a projectile using the upper limbs (Isaac, 1987; Young, 2003). That such tendons would provide advantage in these (often) life-threatening scenarios would submit them to evolutionary pressure, and a longer tendon may thus have been inevitable. Following this, humans would have subsequently also developed efficient movement strategies that best utilise a compliant muscle-tendon unit during slower-speed locomotion (walking and running) in order to retain the ability to function well in rare moments of high-speed locomotion.

Humans, however, are relatively poor sprinters relative to other animals (Carrier, 1984) and perform few high-speed movements in activities of daily living. We also have a lower percentage of fast-twitch fibres in lower-limb propulsive muscles, which provide the highest muscle power outputs, compared to other animals, and even compared to our closest relatives, the chimpanzees (O'Neill et al., 2017; de Diego et al., 2020). Instead, the plantar flexor MTU (and its analogue in many other cursorial species) is more commonly recruited during lower-speed activities such as walking and jogging. Economical walking and running are thought to have provided an important biological advantage over other animals as well as our primate relatives (Carrier, 1984; Liebenberg, 2006; Steudel-Numbers and Wall-Scheffler, 2009), especially as food sources became sparse and travel distances greater from ~8 – 5 mya (Cerling et al., 1997; Cerling et al., 1998). Therefore, a long-standing hypothesis is that hominin transition towards competent bipedalism, and the anatomical and morphological adaptations that allowed it, served to reduce locomotor cost as compared to non-bipedal lineages including those of modern chimpanzees and bonobos (Rodman and Mchenry, 1980).

As human walking is at least as efficient as other animal species (Rubenson et al., 2007), this evolution appears to have been relatively successful.

Important to this 'locomotor cost' hypothesis is that the long, elastic Achilles tendon plays an important energy-storing role that reduces the cost of muscle contraction. Achilles tendon elongation and subsequent recoil during the stance phase of walking and running increases the MTU excursion range relative to muscle length change, reducing the need for muscles to perform mechanical work over large length ranges or at fast shortening velocities, subsequently reducing the energy cost of force production (Lichtwark and Wilson, 2007a; Pontzer et al., 2009; Fletcher et al., 2013; Bohm et al., 2019). During such quasi-isometric contractions, the force-velocity relationship of muscle largely dictates the *maximal* force at any given shortening velocity. However, during walking and running, in which force is generated *submaximally*, shortening velocity dictates the level of activation required to generate a given submaximal force (Stainsby and Lambert, 1979; Fletcher et al., 2013). Reducing the required activation level subsequently reduces the active muscle volume and the considerable energy cost of activation (Stainsby and Lambert, 1979; Chasiotis et al., 1987; Bergstrom and Hultman, 1988). Together, these factors present a model in which the stretch-recoil action of the Achilles tendon optimises the force-velocity profile of the triceps surae muscles to reduce energetic cost for a given locomotor velocity. This phenomenon is demonstrated in Figure 1. Additionally, the isometric force produced by a muscle depends on its sarcomere length, so an optimal muscle length exists (Gordon et al., 1966) and muscle contraction at longer or shorter lengths than optimum results in a lower isometric muscle force (Ramsey and Street, 1940; Gordon et al., 1966). During walking or running, the activation level necessary to produce a given muscle force can be minimised if the muscle operates near optimal length (Figure 2). This optimum may (Rack and Westbury, 1969; Ichinose et al., 1997; Rassier et al., 1999; Holt and Azizi, 2016), or may not (MacDougall et al., 2020), shift to longer lengths with reduced levels of activation. In keeping the level of activation low, muscle energy use is also low. A higher energy cost is seen at shorter than optimal muscle lengths (Hilber et al., 2001) since active force production for a given level of activation is reduced (Hill, 1953; Gordon et al., 1966).

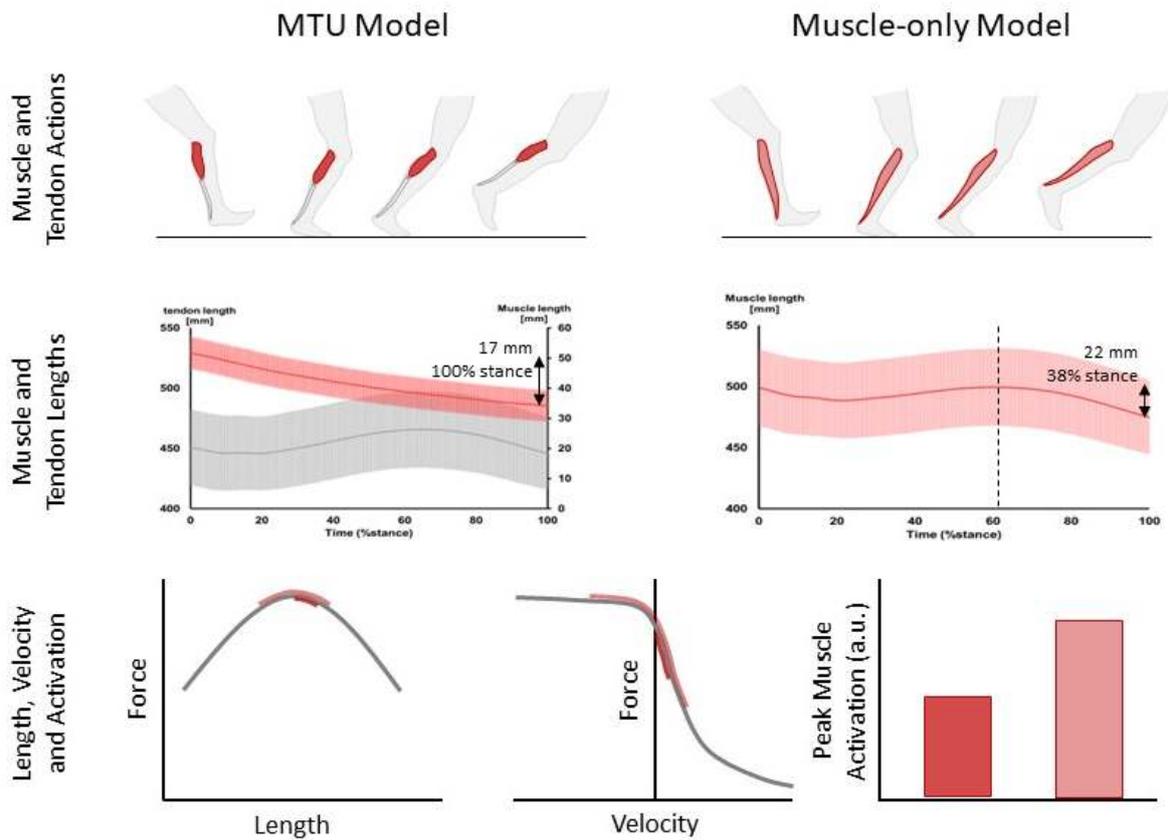

**Figure 1.** MTU model (left column) and muscle-only model (right). Middle row shows separate muscle and tendon length change (MTU model) or muscle length change (muscle-only model) during stance. In the muscle-only model, MTU length change must be accommodated entirely by the muscle, so the magnitude and velocity of shortening is higher. The impact of additional shortening and shortening velocity is shown in the bottom row (force-length relationship, left; force-velocity relationship, middle) and its impact on the required level of activation (right; a.u. arbitrary units).

The tendon operates in the toe region of its recoil force-length relation when muscle forces are relatively low, i.e. it is relatively "slack" at short muscle-tendon unit lengths (Huijing et al., 1989), so muscle shortening is required to stretch the tendon before appreciable force is transmitted and requisite joint moments are produced (Fukutani et al., 2014; Herbert et al., 2015). For a given muscle-tendon unit length, higher forces evoke greater tendon stretch and a corresponding additional muscle shortening. This additional shortening could result in sarcomere length being shorter than optimum or induce shortening-induced force depression, the phenomenon in which active muscle force is depressed during shortening to a given length compared to the isometric force otherwise generated at that same length (Abbott and Aubert, 1952; Edman et al., 1993; Herzog and Leonard, 2000; Rassier and Herzog, 2004). Both factors would necessarily increase the muscle activation level required to reach a given force (Ichinose et al., 1997), thus increasing energy cost, and also increase the rate of activation (i.e. rate of force development) to allow the greater length change in a given time, incurring additional energetic cost (Doke and Kuo, 2007; van der Zee and Kuo, 2021).

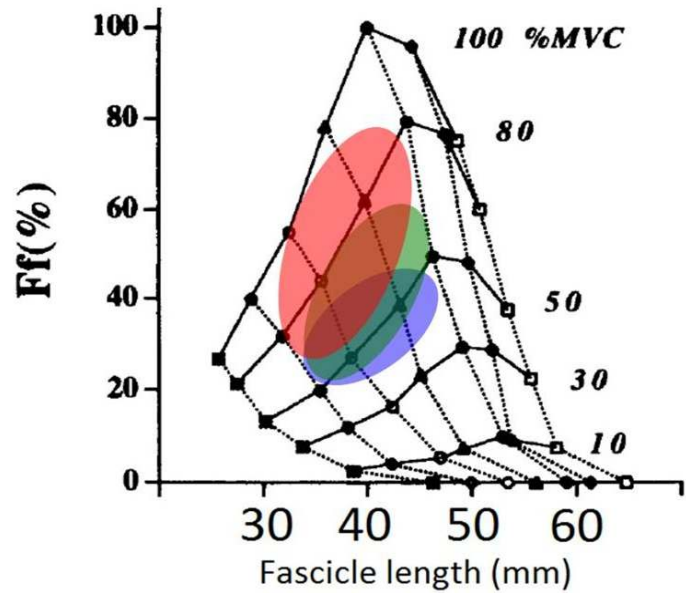

**Figure 2.** Force-fascicle length relation as a function of level of activation (expressed as %MVC). Modified from Ichinose (1997). Blue, green and red circles indicate operating ranges for highly trained males, trained females and trained males, respectively, while running at 75 to 95% of the speed associated with the lactate threshold. Data are from Fletcher et al. 2015.

The capacity to generate the necessary forces in propulsive muscles like the triceps surae is a major factor limiting upright bipedal walking in other primates (Thorpe et al., 2004).

However, the ability to store and release elastic energy in the lower limb is also substantially less in primates than humans and would strongly increase the necessary forces during bipedal locomotion if the relevant force capacity even existed (Sellers et al., 2010). This results directly (but not exclusively) from most primates having short Achilles tendons (Hanna and Schmitt, 2011; Kuo et al., 2013). For example, the Achilles tendon spans ~65% of the length of the plantar flexor-Achilles MTU in humans (Frey, 1913; Prejzner-Morawska and Urbanowicz, 1981) but only ~7.5% in chimpanzees (Rauwerdink, 1991) and <10% in bonobos (Vereecke et al., 2005). While this MTU design, with longer muscle fibres, may assist arboreal locomotion by permitting a greater ankle joint range of motion and for an animal to remain closer to the vertical substrate (DeSilva, 2009) it is clearly detrimental for bipedal locomotion. Instead, the current evidence indicates that long Achilles tendons evolved in the ancestors of modern humans after divergence from the great apes, with recent anatomical evidence that the *Australopithecus Afarensis* (3.9 – 2.9 mya) Achilles tendon possibly spanned ~63% of MTU length (mean of 2 specimens)(McNutt and DeSilva, 2020). *Ipso facto*, the triceps surae muscles, and their constituent fibres, would have been necessarily shorter, reducing the number of active in-series sarcomeres during contraction and thus reducing energetic cost. This evidence is consistent with the theory that the long, in-series, distal lower-limb tendon is an important and unique adaptation in humans over other primates, providing an important functional benefit during activities such as walking.

Nonetheless, contributions from elastic energy storage-return in walking appear to be modest when compared to those at higher locomotor speeds (Lichtwark and Wilson, 2007b; Sawicki et al., 2009). Substantial mechanical energy savings are observed during running (Bramble and Lieberman, 2004; Lichtwark and Wilson, 2007b; Sellers et al., 2010), with mechanical energy savings of ~35% estimated energy storage-return in the Achilles tendon alone (Ker et al., 1987). Humans are unique amongst primates in our capacity for endurance running. The ability to run over long distances with low locomotor cost, e.g. during persistence scavenging and hunting, is speculated to have been of great evolutionary benefit (Bramble and Lieberman, 2004). The benefit of economical running would have further driven the progression towards bipedalism and triggered the anatomical and morphological adaptations required for efficient locomotion. This theory is consistent with the observation of distinct anatomical differences between modern humans and Neanderthals such as a shorter calcaneal tuberosity in humans (Raichlen et al., 2011), indicating a shorter plantar flexor moment arm (Miller and Gross, 1998), that would have increased tendon forces to allow greater elastic energy storage in the Achilles tendon, at least during running (Scholz et al., 2008; Raichlen et al., 2011). Short Achilles moment arms also minimise muscle shortening for a given joint rotation (Nagano and Komura, 2003), reducing the metabolic cost of shortening. Hypothetically, a low locomotor cost may be a key factor influencing the progression towards bipedalism (and particularly running) in humans as well as the simultaneous evolution of distinct anatomical features such as the long Achilles tendon.

**Limitations of, and alternatives to, the 'locomotor cost' hypothesis**

Some limitations exist within this 'locomotor cost' hypothesis. For example:

(i) Some primates have relatively long Achilles tendons and yet are almost completely arboreal and have never been known to demonstrate terrestrial endurance capacity. For example, the Achilles tendon of the gibbon spans 35-45% of muscle-tendon length and is therefore much longer than in the chimpanzee and bonobo (Rauwerdink, 1991).

(ii) Although faster runners may present with better running economy (Fletcher et al., 2009) and the best marathoners (e.g. those approaching sub-2 hours for the marathon) present with exceptional running economy (Lucia et al., 2006b; Jones et al., 2021), running performances and economy are not always strongly associated in elite runners (Mooses et al., 2015). Furthermore, performance times in many endurance running events tend only to be moderately correlated with low locomotor cost; for example, between 25 and 65% of between-athlete 10-km race time variation in well-trained distance runners (Conley and Krahenbuhl, 1980; Powers et al., 1983; Morgan et al., 1989) and <10% of 5-km time variation (Nummela et al., 2006) can be explained by differences in running economy alone. In shorter races (i.e. those above the anaerobic threshold (Svedahl and MacIntosh, 2003)) locomotor cost likely plays a much smaller role in dictating performance compared to maximal oxygen consumption ($\dot{V}O_{2max}$) capacity or fractional utilization of $\dot{V}O_{2max}$. Running economy is also sometimes poorly correlated with performance over the longer marathon and ultramarathon distances (Davies and Thompson, 1979; Sjodin and Svedenhag, 1985; Millet et al., 2011a), and marathon performance is better predicted when $\dot{V}O_{2max}$, running economy, and fractional utilization are considered together (Jones et al. 2021).

(iii) At least in some locomotor tasks, we tend to choose a cadence that is suboptimum from an energetics perspective (e.g. we may transition from walking to running and *vice versa* despite incurring an increased energetic cost (Hreljac, 1993; Tseh et al., 2002)), which in some cases may allow the major propulsive muscles to work at shortening speeds commensurate with peak power production (e.g. in bicycling (Brennan et al., 2018)), or we might choose to optimise gait stability at the expense of energetic cost (e.g. when walking downhill (Hunter et al., 2010)), or choose to locomote under conditions of lower total muscle activation, and presumably sense of effort (crouched walking), rather than conditions requiring higher energetic cost (e.g. walking uphill)(McDonald et al., 2022). That is, we may choose to move under conditions that are energetically more expensive if some other outcome (e.g. power, stability, sense of effort) is optimised, even in prolonged, submaximal exercise.

Of course, a low locomotor cost enables a similar performance to be achieved with a lower $\dot{V}O_{2max}$ (Jones et al., 2021), and low cost should mitigate the rise in both muscle and whole-body temperature, which are key factors influencing the ability to traverse long distances and allowed our ancestors to become diurnal, high-temperature predators (Smil, 2016). However, the focus on locomotor economy ignores the numerous other outcomes that are consequent to the evolution of a long Achilles tendon, including minimisation of both muscular fatigue and sense of effort during locomotion (Carrier, 1984) and the prevention, or minimisation, of muscle damage, such as seen in ultra-marathoning (Millet et al., 2012). These factors will ultimately determine the capacity to continue at a given speed, or the voluntary decision to change locomotor speed (i.e. change the pacing strategy to reduce speed) during locomotion (Marcora et al., 2009; Martin et al., 2018). Additionally, minimisation of limb inertia and transmission reduction of potentially damaging vibrations to bones, muscles and other tissues would be direct and beneficial outcomes of a long Achilles tendon that may theoretically have played an important role in the evolution of bipedalism. Consideration of the importance of these additional factors is needed in order to develop a complete theory as to the role of long distal tendons across cursorial animals, and specifically the evolution of the long Achilles tendon in humans.

**Function of muscle-only vs. muscle-tendon unit (MTU) work production systems**

If a muscle must lengthen and then shorten in a cyclic manner during locomotion, then that muscle (or those muscles) must perform repeated eccentric-concentric work cycles. In a "muscle-only" system, the cost of producing eccentric force (and thus work) during lengthening is less than the cost of isometric force, but the cost of doing concentric work is greater. The total cost is the sum of eccentric and concentric costs. This model approximately reflects that of the chimpanzee (Crompton et al., 2010) and bonobo (Vereecke et al., 2005), and presumedly our last common ancestor, and provides a useful comparative model. The alternative is to place a long elastic tendon in series with the muscle, as in modern humans and referred to here as the muscle-tendon unit, or "MTU", model. In this model the muscle undergoes less length change during locomotion, although it produces force sufficient to drive an oscillating system, and is exemplified by significant tendon lengthening and shortening (Alexander, 1997)(see Figure 1). The muscle must produce brief concentric contractions at discrete points in the cycle to add energy to the system to replace energy that is inevitably lost, but these contractions would occur with minimal length change and at a relatively slow velocity (Lichtwark and Wilson, 2007b; Fletcher et al., 2013; Bohm et al., 2019). In the MTU model, the tendon would dissipate some additional energy, as it displays hysteresis (e.g. ~10%)(Bennett1 et al., 1986), so the total mechanical work done in the system should be slightly greater (and certainly not less) than the total mechanical work in the muscle-only model.

**Effect of muscle-only vs. MTU systems on energetic cost and muscle fatigue**

The total mechanical work done by a muscle when it acts in concert with a tendon to drive locomotion (i.e. the "MTU" model) is always greater than that of a muscle directly powering motion (muscle-only model) because energy is inevitably lost from the tendon during its stretch and recoil in each cycle (energy will be lost from the muscles themselves in both systems, so will be disregarded here for simplicity). However, muscle length change is minimised in the MTU model as tendon length change accounts for a large proportion of the whole MTU length change (Lichtwark et al., 2007)(e.g. Figure 1). The muscle fibres, therefore, shorten less and at a slower velocity for a given MTU length change and velocity (Sawicki et al., 2009), as demonstrated in Figure 3. Since both the greater length change and velocity tend to increase the muscular energy requirement (assuming a constant muscle force)(Woledge et al., 1988; Alexander, 1997; Chow and Darling, 1999), and increases in the rate of force production further add to energetic cost (Doke and Kuo, 2007; van der Zee and Kuo, 2021), the insertion of an in-series tendon can reduce the energetic cost of producing the required MTU work, as discussed above.

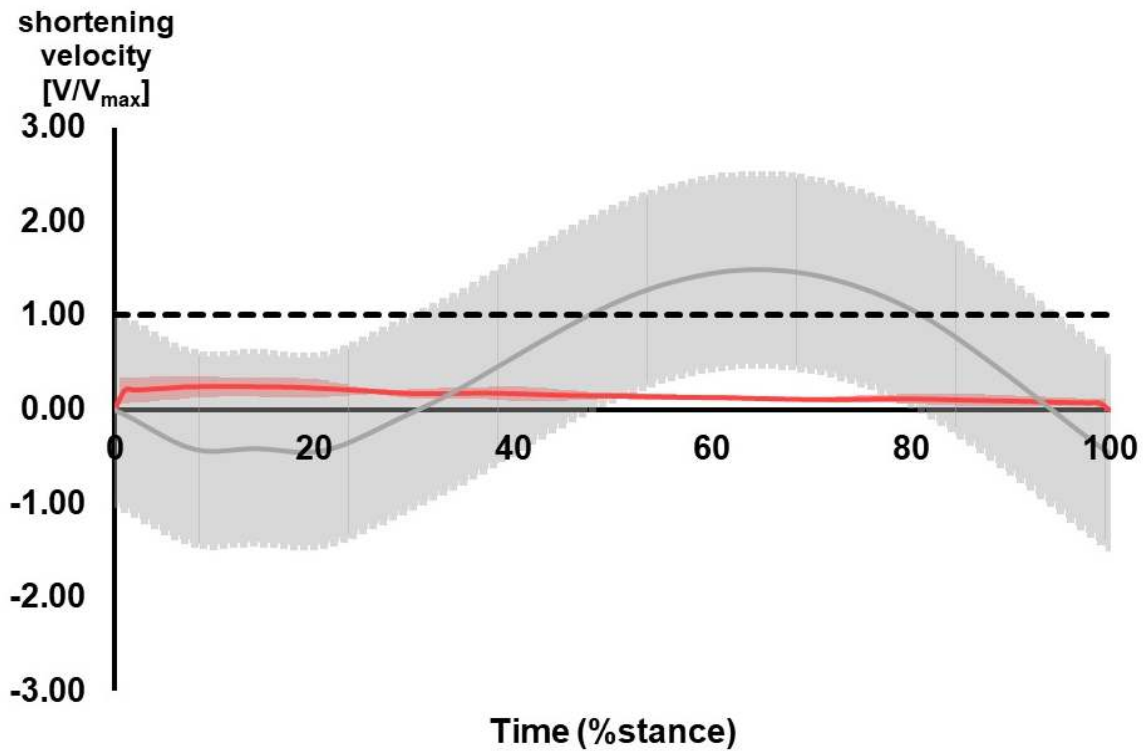

**Figure 3.** Shortening velocity (v) relative to maximal shortening velocity ($v_{max}$) during the stance phase for muscle fascicles in the MTU model (pink) and for the muscle only model (grey). Because in the MTU model, the tendon can accommodate much of the (rapid) lengthening and shortening of the entire MTU, the muscle fascicle shortening velocity is low. In the muscle-only model, any MTU length change must be accommodated solely by the muscle fascicles themselves. Dashed black line indicates maximal shortening velocity of the medial gastrocnemius fascicles (10.8 Lf/s). Note that in the latter portion of the stance phase, muscle fascicles would not be able to shorten fast enough to accommodate the required length change. Fletcher et al. (unpublished observations).

The reduced energetic cost results from the fact that the cost of force generation is not only dependent upon the total muscle work (Fenn, 1923, 1924) but by independent, and partly cumulative, effects of muscle shortening, shortening velocity and, consequently, motor unit recruitment (Roberts et al., 1997; Fletcher et al., 2013). With respect to muscle shortening, concentric muscle work is more costly than isometric force production at a given force level (Hill, 1938; Woledge et al., 1988) yet the force generated at a given level of motor unit recruitment is also smaller, in accordance with the force-velocity relationship (Hill, 1922; Katz, 1939). Therefore, greater muscle activation, and consequently greater motor unit activation, is needed to produce the required force and this is itself associated with an energetic cost (Stainsby and Lambert, 1979). It has been long known that the oxygen requirement of a muscle, which is commensurate with its energy consumption, is more strongly associated with the number of neural pulses provided to the muscle than its shortening or work performed (Fales et al., 1960). In humans, increases in either the force or velocity of shortening in submaximal contractions are predominately influenced by the number of active motor units (Bigland and Lippold, 1954), so the greater the activity of the muscle the greater the energy cost. This is reflected in the strong relationship between oxygen consumption and the integrated electromyogram (EMG) amplitude in human muscles (Bigland-Ritchie and Woods, 1974, 1976). Thus, mechanisms reducing the level of muscle activation will subsequently reduce energetic cost. Since higher muscle forces can be produced during isometric than concentric contractions, a lower muscle activation level is required to meet the force requirements. The smaller proportion of activated muscle mass requires less energy investment to meet force (and work) demands during locomotion. This effect will be greater at faster movement speeds where a muscle operating without an in-series tendon (muscle-only model) would need to produce larger forces at faster shortening speeds to drive motion – at some reasonable running speed, the muscles would not be capable of producing sufficient power to further increase velocity (e.g. see Figure 4). It also follows that an optimum tendon stiffness (in the MTU model) would allow muscle length change to be minimised and for energy cost to be optimised – a very compliant tendon will stretch further under high forces and increase the shortening required by the muscles whilst a very stiff tendon will not stretch appreciably and thus increase the requirement for active muscle length lengthening and then shortening (see Box 1: Is there an "optimum tendon stiffness"?). Muscle energy cost would also be maximised if the tendon was infinitely stiff (i.e. muscle-only model) because it would not change length, so the term 'stiffer' in relation to the tendon is used to represent some optimum stiffness that lies with biological range of tendon stiffness values.

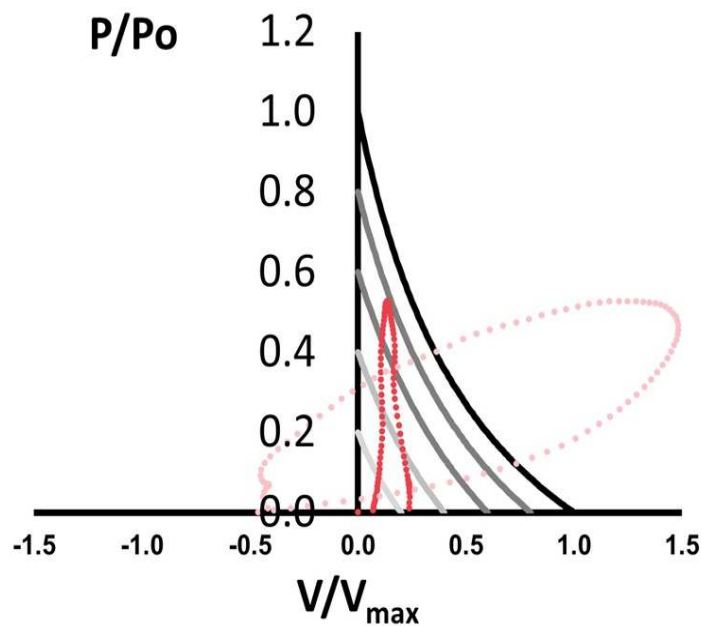

**Figure 4.** Force-velocity curves to demonstrate the operating ranges for muscle fascicle force and shortening velocity in the MTU model (red) and in the muscle-only model (pink). Force-velocity curves are scaled to activation from 20% activation (light grey) to 100% activation (black). In the MTU model (red), muscle fascicles shorten at a velocity that would be predicted from the force-velocity relationship, and maximal estimated activation during stance is 80%. These data align well with the measured and estimated levels of activation of gastrocnemius medialis during running (75-85% maximum activation). In the muscle-only model (pink), the absence of a compliant tendon requires that muscle fascicles shorten rapidly while developing force to support body weight during the stance phase. This shortening is much faster than what would be predicted from the maximal force-velocity curve for skeletal muscle. Therefore, to accommodate this rapid shortening, muscle forces must be drastically reduced; running speed would suffer. Data from Fletcher et al. (unpublished observations).

However, maintaining a quasi-isometric muscle state may also reduce energy consumption according to the force-length relation. Assuming that the muscles work near their optimum length then the energy cost will be low when the contraction is performed near optimal length compared to a contraction at longer or shorter than optimum length in which a greater neural drive is required for the equivalent force. The activation cost is relatively lower (i.e. it represents a lower proportion of the total energetic cost) at optimal length. Evidence for this complex interplay between muscle shortening, shortening velocity, and activation state has been provided for the human plantar flexors (Lichtwark and Barclay, 2010b; Fletcher et al., 2013; Bohm et al., 2019). Based on these arguments, the energy cost of mechanical work during walking can be reduced by the incorporation of a long tendon (of some "optimum" stiffness (Lichtwark and Wilson, 2007b)) into a muscle-tendon unit largely due to its effects on the force-length and force-velocity requirements, and thus activation level, of the muscle. A proportion of the mechanical work during walking can be performed by the tendon, adding to the total mechanical work of the system in which muscle work is minimised. As evidence for this, muscle efficiency during hopping and running tasks is ~40% (Cavagna et al., 1964; Thys et al., 1975)(calculated as mechanical work output per metabolic rate), which is much higher than the ~25% efficiency of muscle's conversion of metabolic to mechanical energy during concentric work (Margaria et al., 1963; Margaria, 1968; Barclay, 2019). That is, movement efficiency is improved, and therefore muscle (metabolic) "fatigue" will be reduced, when tendons are able to return previously stored elastic strain energy during movement. Thus, it makes sense that a key benefit of the MTU system is that it reduces energetic cost of tasks such as locomotion; hence, energetic cost is a primary outcome variable measured in most locomotion studies.

**Broader effects of muscle-only and MTU systems on energetic cost**

Whilst the economic (energy cost) benefits of a long distal tendon are clear and well documented, an important question remains as to whether this is the only, or at least the main, benefit of long, distal tendons.

*Effects on limb inertia*

In addition to the above, locomotor cost is strongly influenced by limb moments of inertia as they swing about their local and remote axes (Willems et al., 1995; Gottschall and Kram, 2005; Umberger, 2010) because this affects the internal work required for limb movement. In human walking, the cost of moving the legs alone may contribute approximately 30% to the total energy cost of walking (Doke et al., 2005; Umberger, 2010) – by contrast, the cost of arm swing is probably very small (Collins et al., 2009) and may even provide a small overall economic advantage (~3%)(Arellano and Kram, 2014) by reducing shoulder and pelvis rotation. The longer muscles required in the muscle-only model will increase distal limb mass and thus limb moment of inertia, and consequently increase energetic cost (Myers and Steudel, 1985). However, insertion of a long tendon reduces distal mass and allows a more proximal mass location. The decreased inertia reduces the cost of moving the limb, which is additional to the economic advantage of elastic mechanisms. The energetic saving from this effect is difficult to determine experimentally, although ~50% greater work is required to swing the lower limb in chimpanzees (with greater distal limb mass) than humans (O'Neill et al., 2022). Also, as detailed in Supplementary File 1 by way of example for an 80-kg male with 1-m leg length and running at 5 m·s$^{-1}$ (18 km·h$^{-1}$), and thus with hip retraction angular velocity of ~5 rad·s$^{-1}$ (287°·s$^{-1}$), the relocation of shank muscle mass 5 cm proximally toward the knee could reduce leg kinetic energy, and thus mechanical work required for limb movement, from 34.1 J to 31.6 J, representing a 7.3% reduction; a more conservative 3-cm shift would reduce energy cost by 4.4%. Assuming that muscle operates concentrically at ~25% efficiency (Margaria, 1968; Barclay, 2019), total muscle work would be reduced by 29.2% and 17.6% for 5-cm and 3-cm mass relocations, respectively. Commensurate with a meaningful inertia-reduction benefit is the smaller size (e.g., cross-sectional area and fibre length) of the plantar flexor muscles in highly trained east African runners who have exceptional running economy (Lucia et al., 2006a; Sano et al., 2015).

The effect of limb inertia is amplified as locomotor speed increases, for example increasing internal work (i.e. kinetic energy) in several species of birds and mammals according to the relation $0.478v^{1.53}$ J·kg$^{-1}$·m$^{-1}$, where $v$ is locomotor speed (m·s$^{-1}$)(Fedak et al., 1982). Based on this relation, internal work increases with speed, and since distal limb segments show larger changes in velocity than proximal segments, reducing distal mass is critical for internal energy cost minimisation (Alexander, 1997, 2002). The evolution of the Achilles tendon would have significantly reduced lower-limb mass, with additional benefit because this mass reduction is provided in the most distal segment. The greater effect of limb inertia reduction at faster locomotor speeds suggests that the adaptation would benefit human running more than walking, consistent with the hypothesis that human running capacity was a key driver of bipedal locomotion in humans.

*Effects on collisional energy loss*

The foot, and therefore the body, collides with the ground at each step causing leg compression and restitution during the ground contact period and providing propulsive force development to propel the body into the next step. This collision, during which the ground reaction force redirects (accelerates) the body's centre of mass (CoM) upward and forward, is an important source of energy loss (Saunders et al., 1953; Donelan et al., 2002; Ruina et al., 2005) as CoM acceleration requires force production and therefore costs metabolic energy. A compliant limb system characterised by bent-knee (crouched) gait could be used to reduce the amplitude of discontinuities in the path of the CoM and thus minimise collisional energy loss (Saunders et al., 1953; Donelan et al., 2002; Ruina et al., 2005; Bertram and Hasaneini, 2013). And this may be easily done in a system driven only by muscles (e.g. the muscle-only model), which would lengthen and shorten to change limb length through the stride cycle to produce a horizontal CoM path. That is, tendons are not explicitly necessary for collision minimisation. However, greater energetic and mechanical costs of transport are observed when a bent-knee gait pattern is adopted than in traditional walking in which the CoM rises and falls slightly within each step (McMahon et al., 1987; Wang et al., 2003; Massaad et al., 2007), despite collisional forces being smaller (McMahon et al., 1987). Part of the additional energy cost comes from the larger external moment arms of force and consequent increases in required joint moments. The associated increased muscular force production would not only increase energetic cost but also accelerate muscle fatigue and body temperature rise (Crompton et al., 1998)(see discussion below), and thus be expected to negatively impact locomotor performance.

Based on the above, neither straight-leg gait (with large collisional energy loss) nor crouched gait (with larger active muscle requirement) are optimal locomotor strategies (Massaad et al., 2007). Instead, a hybrid solution in which a relatively straight-leg gait is used but small-amplitude joint flexions are initiated at the point of foot-ground impact to reduce collision magnitude appears optimal. This gait pattern could be performed using either the muscle-only or MTU models, however some of the collision energy that would be lost during the braking phase of the gait cycle could be stored in the active muscle-tendon units unique to the MTU model. That is, energy can be briefly stored in the tendon and then used later during propulsion rather than being lost to the environment. Thus, collisional energy can be "recycled" to aid propulsion, analogous to a vehicular kinetic energy return system (Boretti, 2013); in this way, the tendon makes an effective kinetic energy return system. Indeed, the propulsion provided by ankle plantar flexion, when appropriately timed, can reduce collisional forces by redirecting the body's CoM just prior to foot-ground contact (Zelik et al., 2014) so a circular argument is created: collision energy can be stored in the Achilles tendon for subsequent propulsion, and the plantar flexion-based propulsion can then help to reduce subsequent collision amplitude and energy loss. As locomotor speed increases and both the ground impact force and requirement for concentric muscle work increase, the re-use of stored collisional energy would become even more useful from an energetic perspective.

**Broader effects of muscle-only and MTU systems on "fatigue"**

An important assumption so far is that the main potential benefit of in-series tendons is their effect on the locomotor cost (through alteration of muscle force production requirements, reduced limb inertia, and store-release of collisional energy), which then influences fatigue based on the intensity-duration relationship. However, an argument can be made that this cost is not the primary determinant of our choice of gait, the locomotor parameters associated with "optimum" gait (e.g. stride length and frequency), or the evolution of long tendons into the human force production machinery (MTU model). Instead, a broader definition of "fatigue" as well as the sense of effort may be needed to understand our capacity to tolerate a given locomotor speed for a given distance, or the decision to adapt movement speed (i.e. alter the pacing strategy) when required. With this consideration, the influence of MTU design on the broader aspects of muscular fatigue and sense of effort should be examined.

*Effects on muscle damage*

Micro-level muscle and connective tissue damage is common during prolonged running (Kim et al., 2007; Millet et al., 2011b; Hoffman et al., 2012). Such damage would need to be avoided in humans running over long distances, especially in hot environments, because of the detrimental effect on running capacity and increased risk of terminal muscle cramping (Hoffman and Stuempfle, 2015; Martinez-Navarro et al., 2020). Limiting the magnitude of concentric-eccentric work cycles (i.e. using an MTU model) reduces both myocellular and connective tissue damage and the resulting disruption of calcium release channels that can occur during unaccustomed, repetitive eccentric muscle contractions (Hyldahl and Hubal, 2014). To minimise muscle damage, and thus to both prolong locomotor activity and ensure that adequate performance in subsequent locomotor bouts (e.g. in the hours or days after the first, damaging bout), reduction of both muscle force and length change during eccentric contractions is required (Penailillo et al., 2015). In a muscle-only model, large eccentric muscle excursions would trigger damage because muscle fibre length change (negative strain) critically affects damage magnitude (Butterfield and Herzog, 2006; Penailillo et al., 2015). However, distal limb muscles tend to remain quasi-isometric during locomotion in an MTU model (Alexander, 2002) so eccentric lengthening is minimised or absent; this system design therefore minimises damage. Using this strategy, highly trained human athletes have been known to complete competitive marathon races (42.1 km) without detectable reductions in muscle function (Petersen et al., 2007), and complete both very long (e.g. 50 – 100 km) and extremely long (>300 km) races with detectable but not critical levels of damage (Fallon et al., 1999; Millet et al., 2011b; Saugy et al., 2013).

One counter argument is that, if the muscle-only model were standard then muscles would perform eccentric contractions on a daily basis, which would confer protection against eccentric exercise-induced muscle damage (i.e. a repeat bout effect would be induced)(Hyldahl et al., 2017). However, it would still be problematic in circumstances in which (i) activity levels are low for at least several days, after which the resumption of daily activities could then trigger significant muscle damage, or (ii) an unaccustomed task is performed for which there is no protection. Thus, an important benefit of the long in-series distal tendon is its capacity to protect the muscle from potentially harmful eccentric contractions, minimising fatigue during exercise and ensuring the ability is retained to perform repeated bouts in the hours and days after an initial bout. This would seem to be a critical, yet often overlooked, evolutionary benefit of the long Achilles tendon in humans and its counterpart in other, particularly cursorial, animals.

*Tendons as vibration dampers*

A collision event is prominent during locomotion even when using a "minimal joint flexion" technique. The higher-frequency components of the ground reaction force during collision at foot-ground initiate tissue vibrations, which would travel rostrally to the head if they remained undamped (Coyles et al., 1999; Wakeling et al., 2002). Active muscles are excellent dampers because strongly bound cross-bridges convert vibrational energy into heat energy for dissipation (Wakeling and Nigg, 2001; Wilson et al., 2001). Muscle activity is therefore required during locomotion for this purpose, i.e., even if muscles were not necessary to drive locomotion, they would be necessary to minimise vibrational energy transmission (Nigg and Liu, 1999; Wakeling et al., 2002). The larger muscles inherent to the muscle-only model might be considered advantageous in this case. However, force production is an inevitable consequence of muscle activation, and this incurs a metabolic cost. So, if the muscle must be active to dampen vibration during the braking phase of foot-ground contact then a strategy in which the energy associated with the muscle work is then stored for subsequent reuse during propulsion might prove useful. Thus, in a system that requires active muscle contraction for vibrational damping, energy storage by an in-series tendon may minimise wasteful energy loss – again, the system functions as an effective energy return system. Evidence indicates that plantarflexor/dorsiflexor co-activation successfully induces fascicle shortening prior to perturbed ground contact in order to protect the muscle from damage from a rapid and forceful lengthening during energy dissipation (Dick et al., 2021). A long, in-series tendon would contribute to some of this additional length change. At faster walking/running speeds where vibrational energy is greater there should be a greater absolute benefit of the MTU model design.

More directly, the tendon itself may also assist with vibration attenuation. The plantar fascia and Achilles tendon form an excellent low-pass filter (Pratt and Williamson, 1995), minimising high frequency vibrational energy transfer from the foot through the proximal upper limb (and head), and subsequently minimising vibration-related damage to muscle (Necking et al., 1992; Necking et al., 1996), tendon (Hansson et al., 1988; Wang et al., 1995) and bone (Carter, 1984). It would also potentially reduce muscle fatigue since less muscle activity would be required for vibrational damping. Correspondingly, reductions in energy cost (through lower muscle activation) would reduce both the rate of muscular fatigue and tissue damage.

*Fibre type use and fatigue*

To generate high power over considerable excursion range, a muscle working in a muscle-only model will need to make use of high-threshold motor units, i.e., faster-twitch fibres. As slow-twitch fibres are more efficient than fast-twitch fibres, at least when muscle speeds are relatively slow (Barclay et al., 1993; Barclay, 1994; He et al., 2000), additional recruitment of faster fibres will be required, which will further increase metabolic cost if shortening velocity is not increased (Woledge, 1998). Because shortening velocity associated with optimal efficiency is increased with progressive motor unit recruitment, energy cost will increase because some of the now active motor units (the newly recruited fast-twitch fibres) will shorten at an energetically inefficient velocity, and fatigue will increase if shortening velocity is not then increased to this optimal velocity. However, the additional mechanical work per unit time would then incur additional metabolic cost. An MTU model circumvents this issue by minimising muscle work and activation, reducing the need for activation of higher-threshold, usually faster-twitch, motor units.

*Effects on muscle and body temperature*

An increase in muscle, and thus whole body, temperature is another source of muscle (and organism) fatigue that would be amplified by the greater metabolic cost of the muscle-only model. Increasing temperatures would be problematic because muscular heat accumulation can reduce net mechanical efficiency of muscular force production (Ferguson et al., 2002; Krustrup et al., 2003) and increase metabolic cost (Brooks et al., 1971; Ferguson et al., 2002), which would reduce exercise endurance time (Edwards et al., 1972; Segal et al., 1986), at least during prolonged bouts. Further, whole-body heat accumulation critically affects exercise capacity, with core (and muscle) temperatures >40°C generally being considered terminal for exercise progression (Gonzalez-Alonso et al., 1999b). The accumulation of heat within muscles and the body depends on the relative rate of heat production to heat dissipation (Taylor et al., 2008). Although there may speculatively be some benefits of an MTU system with regard to heat dissipation (see Supplementary Files 2 and 3), heat production minimisation might be of relatively greater benefit.

Muscles such as those working in the muscle-only model would perform cycles of concentric and eccentric work, both of which generate significant heat through metabolism (concentric; positive work)(Hill, 1949) and the degradation of mechanical energy to heat (eccentric; negative work)(Hill, 1938; Abbott et al., 1951). On the other hand, the use of long in-series tendons reduces muscular work by approximately half during locomotion (or more at faster running speeds)(Cavagna and Kaneko, 1977; Monte et al., 2020) and this would significantly reduce heat production, providing a clear heat-minimisation benefit for prolonged work without critical heat accumulation (Ruxton and Wilkinson, 2011). Thus, for the same external work, greater heat is produced in muscles working according to the muscle-only model than the MTU model.

The balance of heat transfer from the working muscles to neighbouring tissues and the overlying skin versus that is transferred by blood-based convective heat transfer to the core is affected by numerous factors, however when ambient temperature is high (near to or greater than core temperature) >35% of heat production would be directed to the core (Gonzalez-Alonso et al., 1999a). The significant rise in core temperature can rapidly become critical, e.g., in minutes to hours depending on the exercise intensity, metabolic cost, ambient temperature and relative humidity. Furthermore, as heat production through aerobic metabolism exceeds that of anaerobic metabolism (Krustrup et al., 2003), greater heat is generated as a proportion of external work when longer exercise bouts are performed, such as those required for persistence hunting, scavenging, and foraging. Thus, not only would a mechanism by which muscular heat production is minimised provide a significant advantage when ambient temperatures are high, but it provides benefit when prolonged, lower-intensity exercise must be sustained. Thus, the evolution of muscles operating within an MTU model should provide an advantage for longer-duration exercise, especially in warm environments, by minimising both muscular and core temperature increases. This may have been particularly critical during Pliocene (>3 mya; 3°C warmer than today)(Haywood and Valdes, 2004) and other climatic warm periods, as well as for incursions of hominins into warmer latitudes.

It is notable that increases in muscle temperature increase cross-bridge cycling rates, reducing net mechanical efficiency (Ferguson et al., 2002) and subsequently increasing muscular power output during brief or higher-speed/higher-intensity exercise (Bergh and Ekblom, 1979; Sargeant, 1987; Ferretti et al., 1992). So larger muscles that work at higher speeds through larger ranges and with smaller surface area-to-volume ratios might provide an advantage to humans (or other animals) in communities in which such activities predominate over endurance-type behaviours. Thus, whilst the evolution of modern humans in relatively warmer, flat-terrain environments might have benefited from the smaller muscle mass of the MTU model, the possibility cannot be discounted that the migration of humans to colder or mountainous regions may have benefited from larger, less efficient MTU designs.

**Effects on perception of effort**

A fundamental concept so far is that our locomotor capacity is critically linked to physiological and biomechanical phenomena (Rodman and Mchenry, 1980). However, the decision to move is also strongly influenced by our perceptions of both effort and reward; we act (e.g. we move) when the outcome is determined to be subjectively worthwhile (Gendolla and Richter, 2010). When we perceive that a task is more difficult we then have to make decisions as to whether and how to continue, and then exert effort to enact that strategy (Shenhav et al., 2017). Mental effort is laborious and therefore minimised when possible (Kool and Botvinick, 2014), so one might expect that we choose to locomote, or locomote faster, when minimal effort is required yet the reward is great. Mechanisms reducing effort should therefore promote locomotion, and in the much longer term promote the evolution of better movement strategies.

According to the 'sensory tolerance limit' hypothesis (Gandevia, 2001), we may choose to reduce locomotor speed even if the locomotor cost is relatively low in conditions in which movement becomes difficult, including when muscles tire, breathing becomes arduous, pain is felt in exercising or non-exercising muscles, or either ambient or body temperature is high (e.g. St Clair Gibson et al., 2006; Hureau et al., 2018). That is, global feedback in relation to our physiological state is an important limiter of exercise, including locomotor, performance. Global feedback includes that from the working muscles themselves, which strongly influences perception of effort through both feedback and feedforward pathways. In situations of increasing muscle fatigue, defined as a less than anticipated force for a given level of stimulation (MacIntosh and Rassier, 2002), any pain or ill-feeling in the muscle may directly influence our decision to move and require greater effort to do so. However, a greater number of motor units also need to be recruited and firing frequency increases in order to maintain power (i.e. greater effort evokes greater muscle activation). With it, it is theorised, more efferent copies of the motor command are sent from motor to sensory areas of the brain in order to provide a perception of effort associated with the motor output (Duncan et al., 2006; de Morree et al., 2012). Perception of effort therefore increases during fatiguing tasks as a result of elevated efferent input in a feedforward process. As perception of effort increases, the capacity to reach the endpoint of an exercise task may be judged to decrease (in persistence hunting the endpoint may not be defined so the goal may be set to allow exercise for at least several hours, but in many modern sports the endpoint is well known) and power output, i.e., locomotor speed, may be voluntarily reduced in order to reduce effort. Our decision to reduce effort at the expense of increased energetic cost is exemplified by our choice to transition from high-power running gait to low power walking gait at speeds that would increase energetic cost yet reduce sense of effort (Hreljac, 1993; Tseh et al., 2002). Reductions in locomotor speed (or power) or gait transitions reflect a pacing strategy alteration that ultimately dictates our speed (St Clair Gibson et al., 2006). That is, locomotor speed may be influenced not only by locomotor cost or absolute muscle fatigue, as discussed above, but also by the pacing strategy adopted which is established in relation to the cost:reward precept.

Although perception of reward can impact the pacing strategy (motivation, anxiety, stress, hunger/thirst, etc., will influence this)(Lambert et al., 2005), the incorporation of long in-series distal tendons will strongly affect the perception of cost. During a fatiguing locomotor task, an in-series tendon improves the muscle's ability to maintain work output and to reduce the energy cost of contraction because mechanical efficiency is thus improved (Lichtwark and Barclay, 2012). Therefore, both the effort required for muscle contraction and the feedback signals providing information in relation to muscle fatigue (e.g. metabosensitive and pain afferents) will be reduced, so perception of effort is reduced. A human should thus 'feel' more comfortable to locomote at a relatively higher speed for a given distance. According to this hypothesis, a longer Achilles tendon (i.e. the MTU model) increases the likelihood of an individual continuing to move, subsequently increasing food yield or the capacity to migrate to more habitable regions, hence increasing the chance of reproductive success in those individuals and leading to the evolution of a longer Achilles tendon.

However, the reduced metabolic cost of locomotion associated with an MTU model also minimises heat production, and thus reduces the potential negative perceptions of heat accumulation within the body. This increases the likelihood of a human deciding to run longer distances in warmer climates, thus improving long-distance scavenging and hunting practice. Furthermore, if the MTU model is better able to dampen vibration within the musculoskeletal system, then both acute and ongoing benefits may be derived. Acute vibration during submaximal muscle actions tends to increase the perceived effort associated with maintaining a given level of muscle force (Cafarelli and Kostka, 1981; Jones and Hunter, 1985) and negatively affect other psychological states such as perception of comfort during tasks in which vibrations are usually minimal (e.g. bicycling)(Ayachi et al., 2018). Also, at least under some conditions, prolonged vibration reduces subsequent muscle activation capacity and thus limits force production (e.g. 20 min vibration)(Barrera-Curiel et al., 2019), subsequently increasing the voluntary effort required to drive the muscles to maintain force production. In both cases, an increase in voluntary drive to the muscles would increase perception of effort and thus reduce muscular output. In this sense, a long (compliant) distal tendon can both reduce muscle fatigue and the total sensory load projected to the brain, reducing perception of effort and allowing for faster or more prolonged locomotor performance. Given the above, a model in which feedback from numerous systems is influenced by the incorporation of a long, in-series distal tendon, and thus ultimately affects our decision (i.e. capacity) to move, is supported.

**Is there a greater benefit of MTU systems at faster locomotor speeds?**

Numerous benefits to locomotor performance have thus far been described for the MTU model, but some of these benefits are potentially amplified as movement speed increases. With an increase in locomotor speed, at least to speeds ~7 m·s$^{-1}$, the greatest additional work is done by the plantar flexors (Dorn et al., 2012) largely because of the increase in stride length required (for further speed increases the proximal [hip] muscles become increasingly important). Thus, it is reasonable to ask whether there is a specific benefit to placing long, energy-returning tendons in series with distal limb muscles when movement speeds increase.

As the rate of muscle length change increases with locomotor speed in the muscle-only model, the cost of eccentric work will remain relatively constant whilst the cost of concentric work will increase. By contrast, in an MTU system higher muscle forces will be required in order to drive the system at faster speeds (Dean and Kuo, 2011), although the total cost of producing repeated (cyclic) isometric contractions will increase less with movement speed because there is limited need for a greater mechanical work to be performed (see Figures 1 and 3); additional work is done by tendon recoil. Increases in the rate of force development at faster locomotor speeds will also be accomplished by increasing activation rates, which adds its own cost)(Doke and Kuo, 2007; van der Zee and Kuo, 2021), and recruitment of faster-twitch fibres (larger motor units) at lower force levels, i.e., earlier in the force rise phase (Duchateau and Baudry, 2014; Maffiuletti et al., 2016) in both models. This activation increase may thus increase the cost of force production (Bigland-Ritchie and Woods, 1974, 1976; Stainsby and Lambert, 1979; Dean and Kuo, 2011), but is expected to be less than the increasing cost of concentric work and less in the MTU model. Therefore, theoretically, the muscle-only model will become more energetically expensive than the MTU model at higher movement speeds. Based on this consequence, evolutionary pressure for tendon lengthening exerted by the need to locomote at faster speeds (e.g. jog) may be greater than for slower (e.g. walk) locomotor speeds.

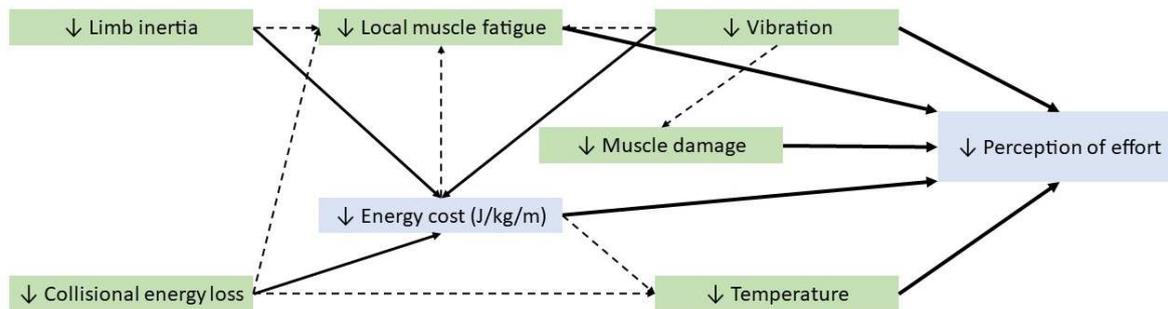

**Figure 5.** The incorporation of a long tendon into the distal muscle-tendon unit (MTU model) will both reduce energy cost, and thus fatigue, as well as decreasing the perception of effort during locomotion (solid arrows in figure). Chief protagonists include reductions in limb inertia, reduced energy loss during foot-ground collisions, storage and reuse of the muscle work required for vibration damping, and reduced muscle damage through both the decreases in muscle length change and vibration propagation (these outcomes may also interact; dashed lines). These benefits should increase both the speed and ease of movement and have greater effect at faster locomotor speeds (i.e. running > walking).

**Conclusions**

Long distal leg tendons such as the Achilles tendon are common in cursorial animals and are known to both amplify muscle power and allow muscle fibres to work over smaller length ranges and at slower shortening speeds, with that energy replaced by stored elastic potential energy. However, little evidence exists that long distal tendons evolved in humans to predominately allow high muscle-tendon power outputs, and indeed we remain relatively powerless compared to many other species. Further, whilst reduced energetic cost of locomotion would be a beneficial adaptive trait, allowing us to walk and run for long distances with minimal energy consumption, other potentially greater benefits exist. As summarised in Figure 5, long distal tendons would reduce muscle fatigue, which is influenced by energetic cost but also (i) reduced limb inertia afforded by the longer tendon and shorter muscles (and corresponding reduced muscle force requirement), (ii) reduced energy loss during foot-ground collisions, (iii) storage and reuse of the muscle work required to dampen the vibrations induced by foot-ground collisions, (iv) reduced exercise-induced muscle damage (especially during eccentric contractions or prolonged vibration), and (v) attenuation of muscle and whole body temperature increase. Many of the above benefits are expected to be greater during running than walking, allowing speculation that running might have been a primary driver of distal muscle-tendon unit morphology in humans. Collectively, these benefits would have strongly influenced perception of effort during locomotor (and other) activities, ultimately allowing a choice to move at faster speeds or for longer. The long distal tendon in modern humans, the Achilles, may therefore be a singular adaptation that provided numerous physiological, biomechanical, and psychological benefits and thus influenced behaviour across multiple tasks, including locomotion. While energy cost may be a variable of interest in locomotor studies, future research should consider the broader range of factors influencing our movement capacity as well as our decision to move over given distances at specific speeds in order to move fully understand the effects of Achilles tendon function, or other factors, on movement performance.

---

**BOX 1: Is there an "optimum tendon stiffness"?**

The most economical (or fastest) walkers (Stenroth et al., 2015) and runners (Arampatzis et al., 2006; Fletcher et al., 2013) have stiffer Achilles tendons and stronger plantar flexor muscles than those who are less economical (or slower)(Fletcher and MacIntosh, 2018). The question may therefore be asked: If we accept that a tendon placed in series with the muscle (the MTU model) is ideal during both slow- and faster-speed locomotion, why might a relatively stiffer tendon (and stronger muscle) be found in those with better walking and running performances? Several possibilities exist.

First, from a mechanical perspective, the efficiency of any oscillating system is optimised when the input energy provided (stretch and recoil of the tendon, assuming the muscle remains quasi-isometric) is temporally matched to the oscillation. So, when the rate of oscillation is fast, the rate of stretch and recoil of the whole muscle-tendon unit, and thus the tendon alone if the muscle remains quasi-isometric, must be fast, and the greater rate is accomplished by increasing stiffness. In humans, 'tuning' of the muscle-tendon interaction at the ankle, for example, in accordance with the movement frequency allows for large energy storage-release of elastic energy (Lichtwark and Wilson, 2005; Takeshita et al., 2006) and increases efficiency above that provided by muscle alone (Biewener and Daley, 2007; Lichtwark and Barclay, 2010a; Dean and Kuo, 2011; Roberts and Azizi, 2011). Additional *ex-vivo* workloop studies in whole muscle-tendon units activated through an intact nerve show that the optimum activation-deactivation rate of the muscle, which elicits the greatest percent contribution to overall MTU power of energy storage and release as well as the highest peak forces, is coupled to the passive stiffness of the muscle-tendon unit (Robertson and Sawicki, 2015). Thus, a runner moving more quickly should require a stiffer tendon (and whole MTU) than a slower runner. That is, as the peak force and locomotor (step) frequency increase, the optimum stiffness of the tendon increases in order for the system to achieve resonance.

Second, the recoil force of a tendon is proportional to its stiffness, approximately according to Hooke's Law ($-F = kx$), so when large ground reaction forces are imposed during faster locomotion a more compliant tendon is less likely to produce recoil force sufficient to optimally accelerate the body (this is separate from the question of whether it would allow the muscle fibres to work quasi-isometrically). Therefore, the muscle would have to shorten more to stretch the tendon sufficiently to transmit the muscular force. Thus, a stiffer tendon would allow for the greater forces required to accelerate the body into each step during high-speed locomotion to be transmitted with less muscle shortening, and thus less work.

Notably, walking and running gaits are associated with specific kinematic constraints, including that the ankle dorsiflexion range of motion has a finite amplitude (<20° dorsiflexion in walking and running)(Mann and Hagy, 1980; Lichtwark et al., 2007). Given such constraints, an MTU can only lengthen a finite distance during ground contact, with lengthening dictated by the MTU force and the tendon stiffness. The average stride cycle force is dictated by body mass (Kram and Taylor, 1990) but maximal tendon force is dictated by running speed, and specifically the duty factor, i.e. the product of ground contact time and stride frequency (Beck et al., 2020). A stiffer tendon would benefit the system because it can store more energy than a compliant tendon for a given total elongation (since, as an approximation, $E = \frac{1}{2} kx^2$, where $E$ is the energy stored, $k$ is the tendon's stiffness, and $x$ is the tendon's elongation, which is a fixed value). This can provide benefit as long as the increase in muscle force required to stretch the stiffer tendon has minimal impact on energetic cost. In actual fact, stiffness would be increased if either the muscle produced more force or the tendon had greater stiffness, and these would need to be matched in some relationship for the system to function properly since a stronger muscle is required to stretch a stiffer tendon. However, stronger muscles require a greater physiological cross-sectional area (Fukunaga et al., 1996), which would increase limb moment of inertia. In the MTU system, a stiffer tendon would be associated with a greater energy storage for a given length change leading to greater potential magnitude and rate of doing positive work during the recoil phase.

The concept that a stiffer tendon can produce a greater recoil force and to do more work for a given length change, enabling greater potential limb accelerations, probably does not completely explain its energetic cost benefit. At faster movement speeds a stiff tendon can contribute force, and thus recoil, while the muscle force is still relatively high. A tendon is stretched when a force is applied. But to recoil, the applied force must decrease (often resulting from the increase in muscle shortening speed, according to the force-velocity relationship, and possibly exacerbated by an increasing internal moment arm during joint extension)(Carrier et al., 1994). For a compliant tendon, the force must be relatively low in order for recoil to occur and for the tendon to contribute mechanical work to the system. Therefore, the force in the muscle-tendon unit during the tendon recoil phase will be necessarily low. However, a stiffer tendon will produce sufficient restoring force to begin its recoil with less decrease in external, and therefore muscle, force, and so it will do work while the muscle force remains high, and the rate of work of the system will be higher than for a relatively more compliant tendon. It is important to note that the optimum tendon stiffness will depend on system mechanics – a tendon that is too stiff will not be of great benefit (see earlier discussion). In many ways this argument is an alternative to the resonance argument above, but clearly articulates the force production advantage of the stiffer system.

An alternative possibility is that the stiffer tendon (and stronger muscle)(Arampatzis et al., 2006; Fletcher and MacIntosh, 2018) may simply be fortuitous for the more economical runner, i.e., it is a requirement rather than a cause. Highly trained runners more commonly use a mid- or forefoot running technique (with whole, or fore-, foot contacting the ground at foot-ground contact) rather than a rearfoot (heel contacting ground) technique (Hasegawa et al., 2007). A greater collision force is generated at ground contact when using a rearfoot technique, increasing energy dissipation because the plantar fascia (of the foot) and Achilles tendon are not optimally loaded at contact (Lieberman et al., 2010). However, some of the collisional energy can be stored in the plantar fascia (Ker et al., 1987) and Achilles tendon (Almonroeder et al., 2013) of mid- or forefoot strikers, which could be reused to help power propulsion. Nonetheless, landing on the forefoot markedly increases Achilles tendon force because of the large external moment arm through which the ground reaction force acts (Lieberman et al., 2010; Almonroeder et al., 2013). Therefore, a compliant Achilles tendon of a forefoot runner would lengthen substantially during foot-ground contact, forcing the ankle into excessive dorsiflexion (i.e. the ankle would work in a manner exceeding its normal constraints) or the muscle fibres to lengthen (and subsequently shorten during propulsion) significantly, unless the muscle produced a large force to cause substantial muscle shortening to stretch the tendon to prevent excessive dorsiflexion; the large concentric excursion would cost metabolic energy. Thus, if a runner were to use a forefoot landing strategy, in which ankle dorsiflexion and thus Achilles tendon loads are higher, then they would need a stiffer tendon and stronger muscle in order to maintain normal, effective gait. In this scenario, the stiffer tendon and stronger plantar flexor muscles could be considered a requirement for forefoot runners, who may also be more economical in their gait at least partly because they can store some of the collisional energy to power propulsion.

**END OF BOX 1**

# REFERENCES

Abbott B, Aubert X, Hill AV. 1951. The absorption of work by a muscle stretched during a single twitch or a short tetanus. Proceedings of the Royal Society of London Series B-Biological Sciences 139:86-104.

Abbott BC, Aubert XM. 1952. The force exerted by active striated muscle during and after change of length. J Physiol 117:77-86.

Alexander RM. 1997. Optimum Muscle Design for Oscillatory Movements. J Theor Biol 184:253-259.

Alexander RM. 2002. Tendon elasticity and muscle function. Comp Biochem Physiol A Mol Integr Physiol 133:1001-1011.

Almonroeder T, Willson JD, Kernozek TW. 2013. The effect of foot strike pattern on achilles tendon load during running. Ann Biomed Eng 41:1758-1766.

Arampatzis A, De Monte G, Karamanidis K, Morey-Klapsing G, Stafilidis S, Bruggemann GP. 2006. Influence of the muscle-tendon unit's mechanical and morphological properties on running economy. J Exp Biol 209:3345-3357.

Arellano CJ, Kram R. 2014. The metabolic cost of human running: is swinging the arms worth it? J Exp Biol 217:2456-2461.

Arnold EM, Hamner SR, Seth A, Millard M, Delp SL. 2013. How muscle fiber lengths and velocities affect muscle force generation as humans walk and run at different speeds. J Exp Biol 216:2150-2160.

Ayachi FS, Drouet JM, Champoux Y, Guastavino C. 2018. Perceptual Thresholds for Vibration Transmitted to Road Cyclists. Hum Factors 60:844-854.

Barclay CJ. 1994. Efficiency of fast- and slow-twitch muscles of the mouse performing cyclic contractions. J Exp Biol 193:65-78.

Barclay CJ. 2019. Efficiency of skeletal muscle. Muscle and exercise physiology:111-127.

Barclay CJ, Constable JK, Gibbs CL. 1993. Energetics of fast- and slow-twitch muscles of the mouse. J Physiol 472:61-80.

Barrera-Curiel A, Colquhoun RJ, Hernandez-Sarabia JA, DeFreitas JM. 2019. The effects of vibration-induced altered stretch reflex sensitivity on maximal motor unit firing properties. J Neurophysiol 121:2215-2221.

Beck ON, Gosyne J, Franz JR, Sawicki GS. 2020. Cyclically producing the same average muscle-tendon force with a smaller duty increases metabolic rate. Proc Biol Sci 287:20200431.

Bennett1 M, Ker1 R, Imery NJ, Alexander1 RM. 1986. Mechanical properties of various mammalian tendons. Journal of Zoology 209:537-548.

Bergh U, Ekblom B. 1979. Influence of muscle temperature on maximal muscle strength and power output in human skeletal muscles. Acta Physiol Scand 107:33-37.

Bergstrom M, Hultman E. 1988. Energy cost and fatigue during intermittent electrical stimulation of human skeletal muscle. J Appl Physiol (1985) 65:1500-1505.

Bertram JE, Hasaneini SJ. 2013. Neglected losses and key costs: tracking the energetics of walking and running. J Exp Biol 216:933-938.

Biewener AA. 1998. Muscle function in vivo: a comparison of muscles used for elastic energy savings versus muscles used to generate mechanical power1. Am Zool 38:703-717.

Biewener AA, Daley MA. 2007. Unsteady locomotion: integrating muscle function with whole body dynamics and neuromuscular control. J Exp Biol 210:2949-2960.

Bigland-Ritchie B, Woods JJ. 1974. Integrated EMG and oxygen uptake during dynamic contractions of human muscles. J Appl Physiol 36:475-479.

Bigland-Ritchie B, Woods JJ. 1976. Integrated electromyogram and oxygen uptake during positive and negative work. J Physiol 260:267-277.

Bigland B, Lippold OC. 1954. Motor unit activity in the voluntary contraction of human muscle. J Physiol 125:322-335.

Blazevich AJ. 2017. Sports Biomechanics - The Basics : Optimising Human Performance, 3rd ed. London: Bloomsbury Publishing PLC.

Bobbert MF, Huijing PA, van Ingen Schenau GJ. 1986. An estimation of power output and work done by the human triceps surae muscle-tendon complex in jumping. J Biomech 19:899-906.
Bohm S, Mersmann F, Santuz A, Arampatzis A. 2019. The force-length-velocity potential of the human soleus muscle is related to the energetic cost of running. Proc Biol Sci 286:20192560.
Boretti A. 2013. Mechanical Hybrid System Comprising a Flywheel and CVT for Motorsport and Mainstream Automotive Applications (2009-01-1312).
Bramble DM, Lieberman DE. 2004. Endurance running and the evolution of Homo. Nature 432:345-352.
Brennan SF, Cresswell AG, Farris DJ, Lichtwark GA. 2018. The effect of cadence on the mechanics and energetics of constant power cycling. Medicine & Science in Sports & Exercise.
Brooks GA, Hittelman KJ, Faulkner JA, Beyer RE. 1971. Temperature, skeletal muscle mitochondrial functions, and oxygen debt. Am J Physiol 220:1053-1059.
Butterfield TA, Herzog W. 2006. Effect of altering starting length and activation timing of muscle on fiber strain and muscle damage. J Appl Physiol (1985) 100:1489-1498.
Cafarelli E, Kostka CE. 1981. Effect of vibration on static force sensation in man. Exp Neurol 74:331-340.
Carrier DR. 1984. The Energetic Paradox of Human Running and Hominid Evolution. Curr Anthropol 25:483-495.
Carrier DR, Heglund NC, Earls KD. 1994. Variable gearing during locomotion in the human musculoskeletal system. Science 265:651-653.
Carter DR. 1984. Mechanical loading histories and cortical bone remodeling. Calcif Tissue Int 36 Suppl 1:S19-24.
Cavagna GA, Kaneko M. 1977. Mechanical work and efficiency in level walking and running. J Physiol 268:467--481.
Cavagna GA, Saibene FP, Margaria R. 1964. Mechanical Work in Running. J Appl Physiol 19:249-256.
Cerling TE, Ehleringer JR, Harris JM. 1998. Carbon dioxide starvation, the development of C-4 ecosystems, and mammalian evolution. Philos T R Soc B 353:159-170.
Cerling TE, Harris JM, MacFadden BJ, Leakey MG, Quade J, Eisenmann V, Ehleringer JR. 1997. Global vegetation change through the Miocene/Pliocene boundary. Nature 389:153-158.
Chasiotis D, Bergstrom M, Hultman E. 1987. ATP utilization and force during intermittent and continuous muscle contractions. J Appl Physiol (1985) 63:167-174.
Chow JW, Darling WG. 1999. The maximum shortening velocity of muscle should be scaled with activation. J Appl Physiol (1985) 86:1025-1031.
Collins SH, Adamczyk PG, Kuo AD. 2009. Dynamic arm swinging in human walking. Proc Biol Sci 276:3679-3688.
Conley DL, Krahenbuhl GS. 1980. Running economy and distance running performance of highly trained athletes. Med Sci Sports Exerc 12:357-360.
Coyles V, Lake M, Lees A. 1999. High frequency movement characteristics of the lower limb during running. Journal of Sports Sciences 17:905-929.
Crompton RH, Sellers WI, Thorpe SK. 2010. Arboreality, terrestriality and bipedalism. Philos Trans R Soc Lond B Biol Sci 365:3301-3314.
Crompton RH, Yu L, Weijie W, Gunther M, Savage R. 1998. The mechanical effectiveness of erect and "bent-hip, bent-knee" bipedal walking in Australopithecus afarensis. J Hum Evol 35:55-74.
Davies CT, Thompson MW. 1979. Aerobic performance of female marathon and male ultramarathon athletes. Eur J Appl Physiol Occup Physiol 41:233-245.
de Diego M, Casado A, Gomez M, Martin J, Pastor JF, Potau JM. 2020. Structural and molecular analysis of elbow flexor muscles in modern humans and common chimpanzees. Zoomorphology 139:277-290.
de Morree HM, Klein C, Marcora SM. 2012. Perception of effort reflects central motor command during movement execution. Psychophysiology 49:1242-1253.

Dean JC, Kuo AD. 2011. Energetic costs of producing muscle work and force in a cyclical human bouncing task. J Appl Physiol (1985) 110:873-880.
DeSilva JM. 2009. Functional morphology of the ankle and the likelihood of climbing in early hominins. Proc Natl Acad Sci U S A 106:6567-6572.
Dick TJM, Clemente CJ, Punith LK, Sawicki GS. 2021. Series elasticity facilitates safe plantar flexor muscle-tendon shock absorption during perturbed human hopping. Proc Biol Sci 288:20210201.
Doke J, Donelan JM, Kuo AD. 2005. Mechanics and energetics of swinging the human leg. J Exp Biol 208:439-445.
Doke J, Kuo AD. 2007. Energetic cost of producing cyclic muscle force, rather than work, to swing the human leg. Journal of Experimental Biology 210:2390-2398.
Donelan JM, Kram R, Kuo AD. 2002. Mechanical work for step-to-step transitions is a major determinant of the metabolic cost of human walking. J Exp Biol 205:3717-3727.
Dorn TW, Schache AG, Pandy MG. 2012. Muscular strategy shift in human running: dependence of running speed on hip and ankle muscle performance. J Exp Biol 215:1944-1956.
Duchateau J, Baudry S. 2014. Insights into the neural control of eccentric contractions. J Appl Physiol (1985) 116:1418-1425.
Duncan MJ, Al-Nakeeb Y, Scurr J. 2006. Perceived exertion is related to muscle activity during leg extension exercise. Research in Sports Medicine 14:179-189.
Edman KA, Caputo C, Lou F. 1993. Depression of tetanic force induced by loaded shortening of frog muscle fibres. J Physiol 466:535-552.
Edwards RH, Harris RC, Hultman E, Kaijser L, Koh D, Nordesjo LO. 1972. Effect of temperature on muscle energy metabolism and endurance during successive isometric contractions, sustained to fatigue, of the quadriceps muscle in man. J Physiol 220:335-352.
Fales JT, Heisey SR, Zierler KL. 1960. Dependency of oxygen consumption of skeletal muscle on number of stimuli during work in the dog. American Journal of Physiology-Legacy Content 198:1333-1342.
Fallon KE, Sivyer G, Sivyer K, Dare A. 1999. The biochemistry of runners in a 1600 km ultramarathon. Br J Sports Med 33:264-269.
Farris DJ, Sawicki GS. 2012. Human medial gastrocnemius force-velocity behavior shifts with locomotion speed and gait. Proc Natl Acad Sci U S A 109:977-982.
Fedak MA, Heglund NC, Taylor CR. 1982. Energetics and mechanics of terrestrial locomotion. II. Kinetic energy changes of the limbs and body as a function of speed and body size in birds and mammals. J Exp Biol 97:23-40.
Fenn WO. 1923. A quantitative comparison between the energy liberated and the work performed by the isolated sartorius muscle of the frog. J Physiol 58:175-203.
Fenn WO. 1924. The relation between the work performed and the energy liberated in muscular contraction. J Physiol 58:373-395.
Ferguson RA, Ball D, Sargeant AJ. 2002. Effect of muscle temperature on rate of oxygen uptake during exercise in humans at different contraction frequencies. J Exp Biol 205:981-987.
Ferretti G, Ishii M, Moia C, Cerretelli P. 1992. Effects of temperature on the maximal instantaneous muscle power of humans. Eur J Appl Physiol Occup Physiol 64:112-116.
Fletcher JR, Esau SP, Macintosh BR. 2009. Economy of running: beyond the measurement of oxygen uptake. J Appl Physiol (1985) 107:1918-1922.
Fletcher JR, MacIntosh BR. 2018. Theoretical considerations for muscle-energy savings during distance running. J Biomech 73:73-79.
Fletcher JR, Pfister TR, Macintosh BR. 2013. Energy cost of running and Achilles tendon stiffness in man and woman trained runners. Physiol Rep 1:e00178.
Frey H. 1913. Der Musculus triceps surae in der Primatenreihe. Morphologische Jahrbuch 47.
Fukunaga T, Roy RR, Shellock FG, Hodgson JA, Edgerton VR. 1996. Specific tension of human plantar flexors and dorsiflexors. J Appl Physiol (1985) 80:158-165.

Fukutani A, Hashizume S, Kusumoto K, Kurihara T. 2014. Influence of neglecting the curved path of the Achilles tendon on Achilles tendon length change at various ranges of motion. Physiol Rep 2.

Galantis A, Woledge RC. 2003. The theoretical limits to the power output of a muscle-tendon complex with inertial and gravitational loads. Proc Biol Sci 270:1493-1498.

Gandevia SC. 2001. Spinal and supraspinal factors in human muscle fatigue. Physiol Rev 81:1725-1789.

Gendolla GH, Richter M. 2010. Effort mobilization when the self is involved: Some lessons from the cardiovascular system. Review of general psychology 14:212-226.

Gonzalez-Alonso J, Calbet JA, Nielsen B. 1999a. Metabolic and thermodynamic responses to dehydration-induced reductions in muscle blood flow in exercising humans. J Physiol 520 Pt 2:577-589.

Gonzalez-Alonso J, Quistorff B, Krustrup P, Bangsbo J, Saltin B. 2000. Heat production in human skeletal muscle at the onset of intense dynamic exercise. J Physiol 524 Pt 2:603-615.

Gonzalez-Alonso J, Teller C, Andersen SL, Jensen FB, Hyldig T, Nielsen B. 1999b. Influence of body temperature on the development of fatigue during prolonged exercise in the heat. J Appl Physiol (1985) 86:1032-1039.

Gordon AM, Huxley AF, Julian FJ. 1966. The variation in isometric tension with sarcomere length in vertebrate muscle fibres. J Physiol 184:170-192.

Gottschall JS, Kram R. 2005. Energy cost and muscular activity required for leg swing during walking. J Appl Physiol (1985) 99:23-30.

Hanna JB, Schmitt D. 2011. Comparative triceps surae morphology in primates: a review. Anatomy Research International 2011.

Hansson HA, Dahlin LB, Lundborg G, Lowenadler B, Paleus S, Skottner A. 1988. Transiently increased insulin-like growth factor I immunoreactivity in tendons after vibration trauma. An immunohistochemical study on rats. Scand J Plast Reconstr Surg Hand Surg 22:1-6.

Hasegawa H, Yamauchi T, Kraemer WJ. 2007. Foot strike patterns of runners at the 15-km point during an elite-level half marathon. J Strength Cond Res 21:888-893.

Haywood AM, Valdes PJ. 2004. Modelling Pliocene warmth: contribution of atmosphere, oceans and cryosphere. Earth and Planetary Science Letters 218:363-377.

He ZH, Bottinelli R, Pellegrino MA, Ferenczi MA, Reggiani C. 2000. ATP consumption and efficiency of human single muscle fibers with different myosin isoform composition. Biophys J 79:945-961.

Hensel H, Bock KD. 1955. Durchblutung und Wärmeleitfähigkeit des menschlichen Muskels. Pflüger's Archiv für die gesamte Physiologie des Menschen und der Tiere 260:361-367.

Herbert RD, Heroux ME, Diong J, Bilston LE, Gandevia SC, Lichtwark GA. 2015. Changes in the length and three-dimensional orientation of muscle fascicles and aponeuroses with passive length changes in human gastrocnemius muscles. J Physiol 593:441-455.

Herzog W, Leonard TR. 2000. The history dependence of force production in mammalian skeletal muscle following stretch-shortening and shortening-stretch cycles. J Biomech 33:531-542.

Hilber K, Sun YB, Irving M. 2001. Effects of sarcomere length and temperature on the rate of ATP utilisation by rabbit psoas muscle fibres. J Physiol 531:771-780.

Hill AV. 1922. The maximum work and mechanical efficiency of human muscles, and their most economical speed. J Physiol 56:19-41.

Hill AV. 1938. The heat of shortening and the dynamic constants of muscle. Proceedings of the Royal Society of London Series B-Biological Sciences 126:136-195.

Hill AV. 1949. Work and heat in a muscle twitch. Proceedings of the Royal Society of London Series B-Biological Sciences 136:220-228.

Hill AV. 1953. The mechanics of active muscle. Proc R Soc Lond B Biol Sci 141:104-117.

Hoffman MD, Ingwerson JL, Rogers IR, Hew-Butler T, Stuempfle KJ. 2012. Increasing creatine kinase concentrations at the 161-km Western States Endurance Run. Wilderness Environ Med 23:56-60.
Hoffman MD, Stuempfle KJ. 2015. Muscle Cramping During a 161-km Ultramarathon: Comparison of Characteristics of Those With and Without Cramping. Sports Med Open 1:24.
Holt NC, Azizi E. 2016. The effect of activation level on muscle function during locomotion: are optimal lengths and velocities always used? Proc Biol Sci 283.
Hreljac A. 1993. Preferred and energetically optimal gait transition speeds in human locomotion. Med Sci Sport Exer 25:1158-1162.
Huijing PA, van Lookeren Campagne AA, Koper JF. 1989. Muscle architecture and fibre characteristics of rat gastrocnemius and semimembranosus muscles during isometric contractions. Acta Anat (Basel) 135:46-52.
Hunter L, Hendrix E, Dean J. 2010. The cost of walking downhill: is the preferred gait energetically optimal? Journal of biomechanics 43:1910-1915.
Hureau TJ, Romer LM, Amann M. 2018. The 'sensory tolerance limit': A hypothetical construct determining exercise performance? Eur J Sport Sci 18:13-24.
Hyldahl RD, Chen TC, Nosaka K. 2017. Mechanisms and Mediators of the Skeletal Muscle Repeated Bout Effect. Exerc Sport Sci Rev 45:24-33.
Hyldahl RD, Hubal MJ. 2014. Lengthening our perspective: morphological, cellular, and molecular responses to eccentric exercise. Muscle Nerve 49:155-170.
Ichinose Y, Kawakami Y, Ito M, Fukunaga T. 1997. Estimation of active force-length characteristics of human vastus lateralis muscle. Acta Anat (Basel) 159:78-83.
Isaac B. 1987. Throwing and human evolution. African Archaeological Review 5:3-17.
Jones AM, Kirby BS, Clark IE, Rice HM, Fulkerson E, Wylie LJ, Wilkerson DP, Vanhatalo A, Wilkins BW. 2021. Physiological demands of running at 2-hour marathon race pace. J Appl Physiol 130:369-379.
Jones LA, Hunter IW. 1985. Effect of muscle tendon vibration on the perception of force. Exp Neurol 87:35-45.
Katz B. 1939. The relation between force and speed in muscular contraction. The Journal of physiology 96:45.
Ker RF, Bennett MB, Bibby SR, Kester RC, Alexander RM. 1987. The Spring in the Arch of the Human Foot. Nature 325:147-149.
Kim HJ, Lee YH, Kim CK. 2007. Biomarkers of muscle and cartilage damage and inflammation during a 200 km run. Eur J Appl Physiol 99:443-447.
Kool W, Botvinick M. 2014. A labor/leisure tradeoff in cognitive control.
Kram R, Taylor CR. 1990. Energetics of running: a new perspective. Nature 346:265-267.
Krustrup P, Ferguson RA, Kjaer M, Bangsbo J. 2003. ATP and heat production in human skeletal muscle during dynamic exercise: higher efficiency of anaerobic than aerobic ATP resynthesis. J Physiol 549:255-269.
Kuo SR, Desilva JM, Devlin MJ, Mcdonald G, Morgan EF. 2013. The Effect of the Achilles Tendon on Trabecular Structure in the Primate Calcaneus. Anat Rec 296:1509-1517.
Lai A, Schache AG, Lin YC, Pandy MG. 2014. Tendon elastic strain energy in the human ankle plantar-flexors and its role with increased running speed. Journal of Experimental Biology 217:3159-3168.
Lambert EV, St Clair Gibson A, Noakes TD. 2005. Complex systems model of fatigue: integrative homoeostatic control of peripheral physiological systems during exercise in humans. Br J Sports Med 39:52-62.
Lichtwark G, Barclay C. 2010a. The influence of tendon compliance on muscle power output and efficiency during cyclic contractions. Journal of Experimental Biology 213:707-714.
Lichtwark GA, Barclay CJ. 2010b. The influence of tendon compliance on muscle power output and efficiency during cyclic contractions. J Exp Biol 213:707-714.

Lichtwark GA, Barclay CJ. 2012. A compliant tendon increases fatigue resistance and net efficiency during fatiguing cyclic contractions of mouse soleus muscle. Acta Physiol (Oxf) 204:533-543.
Lichtwark GA, Bougoulias K, Wilson AM. 2007. Muscle fascicle and series elastic element length changes along the length of the human gastrocnemius during walking and running. J Biomech 40:157-164.
Lichtwark GA, Wilson A. 2005. In vivo mechanical properties of the human Achilles tendon during one-legged hopping. Journal of experimental biology 208:4715-4725.
Lichtwark GA, Wilson AM. 2007a. Is Achilles tendon compliance optimised for maximum muscle efficiency during locomotion? J Biomech 40:1768-1775.
Lichtwark GA, Wilson AM. 2007b. Is Achilles tendon compliance optimised for maximum muscle efficiency during locomotion? Journal of Biomechanics 40:1768-1775.
Liebenberg L. 2006. Persistence hunting by modern hunter-gatherers. Curr Anthropol 47:1017-1025.
Lieberman DE, Bramble DM, Raichlen DA, Shea JJ. 2009. Brains, brawn, and the evolution of human endurance running capabilities. In: The first humans–origin and early evolution of the genus Homo: Springer. p 77-92.
Lieberman DE, Venkadesan M, Werbel WA, Daoud AI, D'Andrea S, Davis IS, Mang'eni RO, Pitsiladis Y. 2010. Foot strike patterns and collision forces in habitually barefoot versus shod runners. Nature 463:531-535.
Lucia A, Esteve-Lanao J, Olivan J, Gomez-Gallego F, San Juan AF, Santiago C, Perez M, Chamorro-Vina C, Foster C. 2006a. Physiological characteristics of the best Eritrean runners-exceptional running economy. Appl Physiol Nutr Metab 31:530-540.
Lucia A, Esteve-Lanao J, Oliván J, Gomez-Gallego F, San Juan AF, Santiago C, Perez M, Chamorro-Vina C, Foster C. 2006b. Physiological characteristics of the best Eritrean runners—exceptional running economy. Applied physiology, nutrition, and metabolism 31:530-540.
MacDougall KB, Devrome AN, Kristensen AM, MacIntosh BR. 2020. Force-frequency relationship during fatiguing contractions of rat medial gastrocnemius muscle. Sci Rep 10:11575.
MacIntosh BR, Rassier DE. 2002. What is fatigue? Can J Appl Physiol 27:42-55.
Maffiuletti NA, Aagaard P, Blazevich AJ, Folland J, Tillin N, Duchateau J. 2016. Rate of force development: physiological and methodological considerations. Eur J Appl Physiol 116:1091-1116.
Mann RA, Hagy J. 1980. Biomechanics of walking, running, and sprinting. Am J Sports Med 8:345-350.
Marcora SM, Staiano W, Manning V. 2009. Mental fatigue impairs physical performance in humans. J Appl Physiol (1985) 106:857-864.
Margaria R. 1968. Positive and negative work performances and their efficiencies in human locomotion. Internationale Zeitschrift für angewandte Physiologie einschließlich Arbeitsphysiologie 25:339-351.
Margaria R, Cerretelli P, Aghemo P, Sassi G. 1963. Energy cost of running. J Appl Physiol 18:367-370.
Martin K, Meeusen R, Thompson KG, Keegan R, Rattray B. 2018. Mental Fatigue Impairs Endurance Performance: A Physiological Explanation. Sports Med 48:2041-2051.
Martinez-Navarro I, Montoya-Vieco A, Collado E, Hernando B, Panizo N, Hernando C. 2020. Muscle Cramping in the Marathon: Dehydration and Electrolyte Depletion vs. Muscle Damage. J Strength Cond Res.
Massaad F, Lejeune TM, Detrembleur C. 2007. The up and down bobbing of human walking: a compromise between muscle work and efficiency. J Physiol 582:789-799.
McDonald KA, Cusumano JP, Hieronymi A, Rubenson J. 2022. Humans trade-off energetic cost with fatigue avoidance while walking. bioRxiv.
McMahon TA, Valiant G, Frederick EC. 1987. Groucho running. J Appl Physiol (1985) 62:2326-2337.
McNutt EJ, DeSilva JM. 2020. Evidence for an elongated Achilles tendon inAustralopithecus. Anat Rec 303:2382-2391.

Miller JA, Gross MM. 1998. Locomotor advantages of Neandertal skeletal morphology at the knee and ankle. J Biomech 31:355-361.
Millet GY, Banfi JC, Kerherve H, Morin JB, Vincent L, Estrade C, Geyssant A, Feasson L. 2011a. Physiological and biological factors associated with a 24 h treadmill ultra-marathon performance. Scand J Med Sci Sports 21:54-61.
Millet GY, Hoffman MD, Morin J-B. 2012. Sacrificing economy to improve running performance—a reality in the ultramarathon? J Appl Physiol 113:507-509.
Millet GY, Tomazin K, Verges S, Vincent C, Bonnefoy R, Boisson RC, Gergele L, Feasson L, Martin V. 2011b. Neuromuscular consequences of an extreme mountain ultra-marathon. Plos One 6:e17059.
Monte A, Maganaris C, Baltzopoulos V, Zamparo P. 2020. The influence of Achilles tendon mechanical behaviour on "apparent" efficiency during running at different speeds. Eur J Appl Physiol 120:2495-2505.
Mooses M, Mooses K, Haile DW, Durussel J, Kaasik P, Pitsiladis YP. 2015. Dissociation between running economy and running performance in elite Kenyan distance runners. Journal of sports sciences 33:136-144.
Morgan DW, Martin PE, Krahenbuhl GS. 1989. Factors affecting running economy. Sports Med 7:310-330.
Myers MJ, Steudel K. 1985. Effect of limb mass and its distribution on the energetic cost of running. J Exp Biol 116:363-373.
Nagano A, Komura T. 2003. Longer moment arm results in smaller joint moment development, power and work outputs in fast motions. J Biomech 36:1675-1681.
Necking LE, Dahlin LB, Friden J, Lundborg G, Lundstrom R, Thornell LE. 1992. Vibration-induced muscle injury. An experimental model and preliminary findings. J Hand Surg Br 17:270-274.
Necking LE, Lundstrom R, Dahlin LB, Lundborg G, Thornell LE, Friden J. 1996. Tissue displacement is a causative factor in vibration-induced muscle injury. J Hand Surg Br 21:753-757.
Nigg BM, Liu W. 1999. The effect of muscle stiffness and damping on simulated impact force peaks during running. J Biomech 32:849-856.
Nummela AT, Paavolainen LM, Sharwood KA, Lambert MI, Noakes TD, Rusko HK. 2006. Neuromuscular factors determining 5 km running performance and running economy in well-trained athletes. Eur J Appl Physiol 97:1-8.
O'Neill MC, Umberger BR, Holowka NB, Larson SG, Reiser PJ. 2017. Chimpanzee super strength and human skeletal muscle evolution. P Natl Acad Sci USA 114:7343-7348.
O'Neill MC, Demes B, Thompson NE, Larson SG, Stern JT, Umberger BR. 2022. Adaptations for bipedal walking: musculoskeletal structure and three-dimensional joint mechanics of humans and bipedal chimpanzees (Pan troglodytes). bioRxiv.
Paluska D, Herr H. 2006. Series elasticity and actuator power output. Ieee Int Conf Robot:1830-1833.
Penaililllo L, Blazevich AJ, Nosaka K. 2015. Muscle fascicle behavior during eccentric cycling and its relation to muscle soreness. Med Sci Sports Exerc 47:708-717.
Petersen K, Hansen CB, Aagaard P, Madsen K. 2007. Muscle mechanical characteristics in fatigue and recovery from a marathon race in highly trained runners. Eur J Appl Physiol 101:385-396.
Pontzer H, Raichlen DA, Sockol MD. 2009. The metabolic cost of walking in humans, chimpanzees, and early hominins. J Hum Evol 56:43-54.
Powers SK, Dodd S, Deason R, Byrd R, Mcknight T. 1983. Ventilatory threshold, running economy and distance running performance of trained athletes. Research Quarterly for Exercise and Sport 54:179-182.
Pratt GA, Williamson MM. 1995. Series elastic actuators. In: Proceedings 1995 IEEE/RSJ International Conference on Intelligent Robots and Systems. Human Robot Interaction and Cooperative Robots: IEEE. p 399-406.
Prejzner-Morawska A, Urbanowicz M. 1981. Morphology of some of the lower limb muscles in primates. In: Primate Evolutionary Biology: Springer. p 60-67.

Rack PM, Westbury DR. 1969. The effects of length and stimulus rate on tension in the isometric cat soleus muscle. J Physiol 204:443-460.

Raichlen DA, Armstrong H, Lieberman DE. 2011. Calcaneus length determines running economy: Implications for endurance running performance in modern humans and Neandertals. J Hum Evol 60:299-308.

Ramsey RW, Street SF. 1940. The isometric length-tension diagram of isolated skeletal muscle fibers of the frog. Journal of Cellular and Comparative Physiology 15:11-34.

Rassier DE, Herzog W. 2004. Effects of shortening on stretch-induced force enhancement in single skeletal muscle fibers. Journal of Biomechanics 37:1305-1312.

Rassier DE, MacIntosh BR, Herzog W. 1999. Length dependence of active force production in skeletal muscle. J Appl Physiol (1985) 86:1445-1457.

Rauwerdink IG. 1991. Muscle fibre and tendon lengths in the distal limb segments of primates. Z Morphol Anthropol 78:331-340.

Roberts TJ. 2016. Contribution of elastic tissues to the mechanics and energetics of muscle function during movement. J Exp Biol 219:266-275.

Roberts TJ, Azizi E. 2011. Flexible mechanisms: the diverse roles of biological springs in vertebrate movement. Journal of experimental biology 214:353-361.

Roberts TJ, Marsh RL, Weyand PG, Taylor CR. 1997. Muscular force in running turkeys: the economy of minimizing work. Science 275:1113-1115.

Robertson BD, Sawicki GS. 2015. Unconstrained muscle-tendon workloops indicate resonance tuning as a mechanism for elastic limb behavior during terrestrial locomotion. Proc Natl Acad Sci U S A 112:E5891-5898.

Rodman PS, Mchenry HM. 1980. Bioenergetics and the Origin of Hominid Bipedalism. Am J Phys Anthropol 52:103-106.

Rubenson J, Lloyd DG, Besier TF, Heliams DB, Fournier PA. 2007. Running in ostriches (Struthio camelus): three-dimensional joint axes alignment and joint kinematics. J Exp Biol 210:2548-2562.

Ruina A, Bertram JE, Srinivasan M. 2005. A collisional model of the energetic cost of support work qualitatively explains leg sequencing in walking and galloping, pseudo-elastic leg behavior in running and the walk-to-run transition. J Theor Biol 237:170-192.

Ruxton GD, Wilkinson DM. 2011. Thermoregulation and endurance running in extinct hominins: Wheeler's models revisited. J Hum Evol 61:169-175.

Sano K, Nicol C, Akiyama M, Kunimasa Y, Oda T, Ito A, Locatelli E, Komi PV, Ishikawa M. 2015. Can measures of muscle-tendon interaction improve our understanding of the superiority of Kenyan endurance runners? Eur J Appl Physiol 115:849-859.

Sargeant AJ. 1987. Effect of muscle temperature on leg extension force and short-term power output in humans. Eur J Appl Physiol Occup Physiol 56:693-698.

Saugy J, Place N, Millet GY, Degache F, Schena F, Millet GP. 2013. Alterations of Neuromuscular Function after the World's Most Challenging Mountain Ultra-Marathon. Plos One 8:e65596.

Saunders JB, Inman VT, Eberhart HD. 1953. The major determinants in normal and pathological gait. J Bone Joint Surg Am 35-A:543-558.

Sawicki GS, Lewis CL, Ferris DP. 2009. It Pays to Have a Spring in Your Step. Exercise and Sport Sciences Reviews 37:130-138.

Sawicki GS, Sheppard P, Roberts TJ. 2015. Power amplification in an isolated muscle-tendon unit is load dependent. J Exp Biol 218:3700-3709.

Scholz MN, Bobbert MF, van Soest AJ, Clark JR, van Heerden J. 2008. Running biomechanics: shorter heels, better economy. Journal of Experimental Biology 211:3266-3271.

Segal SS, Faulkner JA, White TP. 1986. Skeletal muscle fatigue in vitro is temperature dependent. J Appl Physiol (1985) 61:660-665.

Sellers WI, Pataky TC, Caravaggi P, Crompton RH. 2010. Evolutionary Robotic Approaches in Primate Gait Analysis. Int J Primatol 31:321-338.

Shenhav A, Musslick S, Lieder F, Kool W, Griffiths TL, Cohen JD, Botvinick MM. 2017. Toward a rational and mechanistic account of mental effort. Annual review of neuroscience 40:99-124.
Sherwood L. Human physiology : from cells to systems, Ninth edition. ed.
Sjodin B, Svedenhag J. 1985. Applied physiology of marathon running. Sports Med 2:83-99.
Smil V. 2016. Running, sweating, and persistence hunting [Numbers Don't Lie]. IEEE Spectrum 53:26-26.
St Clair Gibson A, Lambert EV, Rauch LH, Tucker R, Baden DA, Foster C, Noakes TD. 2006. The role of information processing between the brain and peripheral physiological systems in pacing and perception of effort. Sports Med 36:705-722.
Stainsby WN, Lambert CR. 1979. Determination of oxygen uptake in skeletal muscle. Exerc Sport Sci Rev 7:125-151.
Stenroth L, Sillanpaa E, McPhee JS, Narici MV, Gapeyeva H, Paasuke M, Barnouin Y, Hogrel JY, Butler-Browne G, Bijlsma A, Meskers CG, Maier AB, Finni T, Sipila S. 2015. Plantarflexor Muscle-Tendon Properties are Associated With Mobility in Healthy Older Adults. J Gerontol A Biol Sci Med Sci 70:996-1002.
Steudel-Numbers KL, Wall-Scheffler CM. 2009. Optimal running speed and the evolution of hominin hunting strategies. J Hum Evol 56:355-360.
Svedahl K, MacIntosh BR. 2003. Anaerobic threshold: the concept and methods of measurement. Canadian journal of applied physiology 28:299-323.
Takeshita D, Shibayama A, Muraoka T, Muramatsu T, Nagano A, Fukunaga T, Fukashiro S. 2006. Resonance in the human medial gastrocnemius muscle during cyclic ankle bending exercise. J Appl Physiol 101:111-118.
Taylor N, Kondo N, Kenney WL. 2008. The physiology of acute heat exposure, with implications for human performance in the heat.
Thorpe CT, Godinho MSC, Riley GP, Birch HL, Clegg PD, Screen HRC. 2015. The interfascicular matrix enables fascicle sliding and recovery in tendon, and behaves more elastically in energy storing tendons. J Mech Behav Biomed Mater 52:85-94.
Thorpe SK, Crompton RH, Wang WJ. 2004. Stresses exerted in the hindlimb muscles of common chimpanzees (Pan troglodytes) during bipedal locomotion. Folia Primatol (Basel) 75:253-265.
Thys H, Cavagna GA, Margaria R. 1975. The role played by elasticity in an exercise involving movements of small amplitude. Pflugers Arch 354:281-286.
Tseh W, Bennett J, Caputo JL, Morgan DW. 2002. Comparison between preferred and energetically optimal transition speeds in adolescents. Eur J Appl Physiol 88:117-121.
Umberger BR. 2010. Stance and swing phase costs in human walking. J R Soc Interface 7:1329-1340.
van der Zee TJ, Kuo AD. 2021. The high energetic cost of rapid force development in muscle. Journal of Experimental Biology 224.
Vereecke EE, D'Aout K, Payne R, Aerts P. 2005. Functional analysis of the foot and ankle myology of gibbons and bonobos. J Anat 206:453-476.
Wakeling JM, Nigg BM. 2001. Modification of soft tissue vibrations in the leg by muscular activity. J Appl Physiol (1985) 90:412-420.
Wakeling JM, Pascual SA, Nigg BM. 2002. Altering muscle activity in the lower extremities by running with different shoes. Med Sci Sports Exerc 34:1529-1532.
Wang WJ, Crompton RH, Li Y, Gunther MM. 2003. Energy transformation during erect and 'bent-hip, bent-knee' walking by humans with implications for the evolution of bipedalism. J Hum Evol 44:563-579.
Wang XT, Ker RF, Alexander RM. 1995. Fatigue rupture of wallaby tail tendons. J Exp Biol 198:847-852.
Willems PA, Cavagna GA, Heglund NC. 1995. External, internal and total work in human locomotion. J Exp Biol 198:379-393.
Wilson AM, McGuigan MP, Su A, van Den Bogert AJ. 2001. Horses damp the spring in their step. Nature 414:895-899.

Woledge RC. 1998. Possible effects of fatigue on muscle efficiency. Acta Physiol Scand 162:267-273.

Woledge RC, Wilson MG, Howarth JV, Elzinga G, Kometani K. 1988. The energetics of work and heat production by single muscle fibres from the frog. Adv Exp Med Biol 226:677-688.

Young RW. 2003. Evolution of the human hand: the role of throwing and clubbing. J Anat 202:165-174.

Zelik KE, Huang TW, Adamczyk PG, Kuo AD. 2014. The role of series ankle elasticity in bipedal walking. J Theor Biol 346:75-85.

# SUPPLEMENTARY FILE 1 – Effect of shank mass distribution on kinetic energy of the leg during running

Leg swing during walking, and particularly running, costs energy. The kinetic energy of a limb can be calculated as the sum of kinetic energies of the limb segments swinging about their local and remote axes. In one 80-kg runner with leg length of 1 m, the segment masses and centre of mass locations were estimated during running using the data of Clauser et al. (Sherwood), as shown in Blazevich (Blazevich, 2017). Distances from the hip centre of rotation to centre of mass locations of the foot, shank and thigh were 0.90 m, 0.60 m and 0.25 m, respectively, and segment masses were 1.20 kg, 3.44 kg and 8.24 kg. The kinetic energy ($E_K$) of the retracting (late swing-stance) leg at an angular velocity ($\omega$) immediately prior to foot-ground contact of 5 rad·s$^{-1}$ was calculated as $\Sigma(E_{K,foot} + E_{K,shank} + E_{k,thigh})$, with $E_K$ equal to $\frac{1}{2}(mk^2)\cdot\omega^2$, where *m* is segment mass and *k* is its radius of gyration.

Local moments of inertia were 0.0038, 0.0504 and 0.1052 kg·m$^2$ for foot, shank and thigh segments, respectively, so $E_K$ about the local axes for the foot [$\frac{1}{2}(0.0038)\times5^2$], shank [$\frac{1}{2}(0.0504)\times5^2$] and thigh [$\frac{1}{2}(0.1052)\times5^2$] were 0.048, 0.63 and 1.32 J and the total local $E_K$ was 1.99 J. As shown below, $E_K$ about local axes represented ~5% of the total $E_K$ and would be predicted to change by only a small fraction if mass was redistributed in the shank only. It was therefore removed from calculations, and only $E_K$ about the remote axis was investigated.

Remote moments of inertia were 1.20, 3.44 and 8.24 kg·m$^2$ for foot, shank and thigh segments, respectively, so $E_K$ about the remote axes for the foot [$\frac{1}{2}(1.20\times0.9^2)\times5^2$], shank [$\frac{1}{2}(3.44\times0.6^2)\times5^2$] and thigh [$\frac{1}{2}(8.24\times0.25^2)\times5^2$] were 12.2, 15.5 and 6.44 J, and the total local $E_K$ was 34.1 J.

If the mass of the shank segment only were moved 0.05 m proximally then the remote $E_K$ would equal $\frac{1}{2}(3.44 \times 0.55^2) \times 5^2 = 13.0$ J, reducing $E_K$ by 2.5 J, or 7.3%. A more modest mass relocation of 0.03 m would result in a 4.4% reduction in $E_K$.

During leg protraction (recovery) in a well-trained runner the knee is flexed acutely, allowing the shank and foot to travel closer to the hip axis of rotation and with the longitudinal axis of the shank approximately perpendicular to the line joining the shank centre of mass location and the hip joint. Thus, moving the shank mass proximally should have no appreciable effect on limb protraction kinetic energy, and thus energy cost of running.

**SUPPLEMENTARY FILE 2 – Possible effects of muscle model on heat dissipation**

Movement of heat to the environment may be achieved through blood flow-mediated convection to the core or skin, or directly through conduction to neighbouring tissues and then the skin. In some cases, more than half of the metabolic heat developed during prolonged exercise in the heat may be dissipated directly to the environment surrounding the working musculature, including the skin (Gonzalez-Alonso et al., 1999b). Thus, in addition to increased blood flow, an increase in muscle surface area relative to volume might speculatively increase this capacity. A muscle that produces force over a large excursion range (as in the muscle-only model) will necessarily be long (Biewener, 1998), and yet still require a minimum cross-sectional area to produce the requisite forces for locomotion. The volume of muscles operating in the muscle-only model must therefore be greater than in the MTU model. In the example shown in detail in Supplementary File 3, a cylindrical muscle with length 40 cm (≈ male human shank) and mean diameter 6 cm would have a surface area-to-volume ratio of 0.717 whereas a muscle working within an MTU system with similar cross-section but 25% of the length (see supplementary file for justification) would have a ratio of 0.866, or ~21% greater. This might speculatively provide an advantage for heat dissipation to surrounding tissues and the skin during prolonged physical activity, particularly in hot ambient temperatures (Ruxton and Wilkinson, 2011).

However, the important question is whether the greater ratio would meaningfully assist transductive heat dissipation into the local muscular environment and subsequently the surrounding body for removal (through radiation, sweating or breathing), ensuring that muscle (and whole body) temperatures remain lower and heat accumulation is less likely to contribute to fatigue (Lieberman et al., 2009). In fact, the potential effect of the muscle's surface area-to-volume ratio must be set against the relative importance of direct heat transfer to that provided by blood perfusion, especially since heat transfer through tissues within the body or to the environment is slow (Hensel and Bock, 1955). Venous blood temperature increases immediately upon commencement of intense exercise, suggesting that it rapidly removes heat energy from working muscles (Gonzalez-Alonso et al., 2000) and heat transfer to the blood is the primary mechanism of heat removal from muscles. The precise quantity of heat liberated directly to the skin is difficult to measure accurately *in vivo* and differs depending on factors such as the muscle-skin and muscle-environment temperature differences, effects of convection, rate of heat production per vascular flux, etc., however it is likely to be relatively small. One might consider, then, that even a ~20% difference in surface area-to-volume might produce only a small thermoregulatory benefit through direct heat transfer to the skin for removal. In itself, this may not be a primary advantage of the MTU model. Therefore, long in-series tendons in distal muscle groups should significantly reduce muscle heat production during locomotion and in turn reduce whole-body heat accumulation, but the effect of reducing the surface area-to-volume ratio on heat accumulation itself might only be small.

**SUPPLEMENTARY FILE 3 – Effect of muscle model on surface area-to-volume ratio**

The volume of muscles operating in the muscle-only model must be greater than in the MTU model, and this speculatively reduces surface heat loss in proportion to the difference in surface area; i.e. direct heat exchange should be proportional to the surface area-to-volume ratio, if heat can indeed be well transferred.

To estimate the effects of muscle design on surface area-to-volume, we could take a cylindrical muscle with length 40 cm (≈ male human shank) and mean diameter 6 cm, which would thus have a surface area-to-volume ratio of 0.717 (811 $cm^2$/1131 $cm^3$). However, a muscle working within an MTU produces force over a smaller excursion range and may thus be shorter. Whilst the muscle might have a greater peak force requirement, a larger cross-sectional area may not be necessary since muscles produce greater forces under quasi-isometric conditions than under dynamic conditions in accordance with the force-velocity relationship. Thus, the cross-sectional area-to-peak force ratio can be smaller. If the muscle in the MTU model was 25% of the length of that in the muscle-only model but had the same cross-sectional area then its surface area-to-volume ratio would be 0.866 (245 $cm^2$/283 $cm^3$), which is ~21% greater than the muscle-only model.

This ~20% difference in surface area-to-volume ratio should speculatively provide an advantage for heat dissipation to surrounding tissues and the skin during prolonged physical activity, particularly in hot ambient temperatures (Ruxton and Wilkinson, 2011). However, this is predicated on direct heat transfer being substantial *in vivo* (see main text for further discussion).